\documentclass[twocolumn]{aastex61}

\def\mJB{{\rm mJy~beam^{-1}}}

\def\mJ{{\rm mJy}}
\def\kms{{\rm km~s^{-1}}}
\def\Ms{M_{\sun}}
\def\Ls{L_{\sun}}

\def\II{{\rm I\hspace{-0.1em}I}}

\received{}
\revised{\today}
\accepted{}
\submitjournal{ApJ}

\shorttitle{SMM4 in Serpens Main}
\shortauthors{Aso et al.}

\begin{document}

\title{The Distinct Evolutionary Nature of two Class 0 Protostars in Serpens Main SMM4}

\correspondingauthor{Yusuke Aso}
\email{yaso@asiaa.sinica.edu.tw}

\author[0000-0002-8238-7709]{Yusuke Aso}
\affil{Academia Sinica Institute of Astronomy and Astrophysics P.O. Box 23-141, Taipei 106, Taiwan}

\author[0000-0001-9304-7884]{Naomi Hirano}
\affil{Academia Sinica Institute of Astronomy and Astrophysics P.O. Box 23-141, Taipei 106, Taiwan}

\author[0000-0003-3283-6884]{Yuri Aikawa}
\affil{Department of Astronomy, Graduate School of Science, The University of Tokyo, 7-3-1 Hongo, Bunkyo-ku, Tokyo 113-0033, Japan}

\author[0000-0002-0963-0872]{Masahiro N. Machida}
\affil{Department of Earth and Planetary Sciences, Faculty of Sciences Kyushu University, Fukuoka 812-8581, Japan}

\author[0000-0003-0845-128X]{Shigehisa Takakuwa}
\affil{Academia Sinica Institute of Astronomy and Astrophysics P.O. Box 23-141, Taipei 106, Taiwan}
\affil{Department of Physics and Astronomy, Graduate School of Science and Engineering, Kagoshima University, 1-21-35 Korimoto, Kagoshima, Kagoshima 890-0065, Japan}

\author[0000-0003-1412-893X]{Hsi-Wei Yen}
\affil{European Southern Observatory, Karl-Schwarzschild-Str. 2, D-85748 Garching, Germany}

\author[0000-0001-5058-695X]{Jonathan P. Williams}
\affil{Institute for Astronomy, University of Hawaii at Manoa, Honolulu, Hawaii, USA}



\begin{abstract}
We have observed the submillimeter continuum condensation SMM4 in Serpens Main using the Atacama Large Millimeter/submillimeter Array (ALMA) during its Cycle 3 in 1.3 mm continuum, $^{12}$CO $J=2-1$, SO $J_N=6_5-5_4$, and C$^{18}$O $J=2-1$ lines at angular resolutions of $\sim 0\farcs 55$ (240 au). The 1.3 mm continuum emission shows that SMM4 is spatially resolved into two protostars embedded in the same core: SMM4A showing a high brightness temperature, 18 K,  with little extended structure and SMM4B showing a low brightness temperature, 2 K, with compact and extended structures. Their separation is $\sim 2100$ au. Analysis of the continuum visibilities reveals a disk-like structure with a sharp edge at $r\sim240$ au in SMM4A, and a compact component with a radius of 56 au in SMM4B. 
The $^{12}$CO emission traces fan-shaped and collimated outflows associated with SMM4A and SMM4B, respectively. The blue and red lobes of the SMM4B outflow have different position angles by $\sim 30\arcdeg$. Their inclination and bending angles in the 3D space are estimated at $i_b\sim 36\arcdeg$, $i_r\sim 70\arcdeg$, and $\alpha\sim 40\arcdeg$, respectively. The SO emission traces shocked regions, such as cavity walls of outflows and the vicinity of SMM4B. The C$^{18}$O emission mainly traces an infalling and rotating envelope around SMM4B. The C$^{18}$O fractional abundance in SMM4B is $\sim 50$ times smaller than that of the interstellar medium. These results suggest that SMM4A is more evolved than SMM4B. Our studies in Serpens Main demonstrate that continuum and line observations at millimeter wavelengths allow us to differentiate evolutionary phases of protostars within the Class 0 phase.
\end{abstract}




\section{INTRODUCTION} \label{sec:intro}
Class 0, the earliest evolutionary phase of protostars is defined as young stellar objects (YSOs) detected at (sub)millimeter wavelengths but not detected at near- and mid-infrared wavelengths, 2-25 $\micron$ \citep{an1993}. While the Class 0 phase is thought to have a shorter lifetime, $\lesssim 0.3$ Myr \citep{an.mo1994,ev2009,du2015}, than the other phases, the phase is filled with various dynamical and chemical phenomena. Because Class 0 protostars cannot be directly observed at infrared wavelengths, millimeter/submillimeter observations have been developed to investigate them, particularly, interferometric observations at high spatial resolution. These observational studies suggest that circumstellar disks start to form in the Class 0 phase \citep[e.g.,][]{to2012,mu.la2013,oh2014,ye2017,aso2017a}, and reveal morphology and kinematics of molecular outflows/jets in Class 0 protostars \citep[e.g.,][]{hi2006,hi2010,ye2017,le2017,aso2017b}. Opening angles of protostellar outflows widen on a timescale similar to the lifetime of Class 0 protostars \citep{ar.sa2006,ma.ho2013} and molecular abundances are predicted, and observed, to vary within the Class 0 phase \citep{ha2015,ai2012,hi.li2014,aso2017b}. These studies suggest that we can distinguish finer levels of evolution within the Class 0 phase than the simple SED classification from optical/infrared observations.

In order to study Class 0 evolutionary phases, we observed the star forming cluster Serpens Main with the Atacama Large Millimeter/Submillimeter Array (ALMA) in $^{12}$CO $J=2-1$ (230.538 GHz), C$^{18}$O $J=2-1$ (219.560 GHz), and SO $J_N=6_5-5_4$ (219.949 GHz) lines and 1.3 mm continuum emission. The distance to Serpens Main is 429 pc \citep{dz2011}. Our targets are the submillimeter condensations identified with James Clerk Maxwell Telescope (JCMT) observations and one of these condensations, SMM4, are focused in this paper.

The distribution of YSOs in Serpens Main, including six Class 0 protostars \citep{du2008}, suggests that the star forming cluster experienced two episodes of star formation and the latter occurred 0.5 Myr ago \citep{ha2007,ka2004}, possibly due to a collision of two clouds or flows occurring around the SE subcluster \citep{d-c2010,d-c2011}.
SMM4 is located near the western edge of the SE subcluster in Serpens Main. \citet{leKI2014} observed Serpens Main using Combined Array for Research in Millimeter-wave Astronomy (CARMA) in 3 mm continuum emission at a $\sim 7\farcs5$ angular resolution and revealed that SMM4 has the highest brightness in the SE subcluster. They estimated its deconvolved size and mass to be 2200 au and $5.9\ \Ms$, respectively, assuming a dust temperature of 20 K and a dust opacity of $\kappa (3\ {\rm mm})=0.0027\ {\rm cm}^{2}~{\rm g}^{-1}$. They also detected HCN $(1-0)$ line emission ($n_{\rm cr}\sim 1\times 10^6\ {\rm cm}^{-3}$) stronger in the north and northwest of SMM4 than in other regions of the SE subcluster.

Our ALMA observations and data reduction are described in Section \ref{sec:observations}. In Section \ref{sec:results}, we present results of the continuum and emission lines derived from the ALMA observations in SMM4, as well as its SED. Further analyses will be performed in Section \ref{sec:analysis} to investigate configurations of a bipolar $^{12}$CO outflow, visibilities of dust continuum, and C$^{18}$O abundance. A possible origin of a bending outflow and evolutionary phases of two protostars discovered in this region, SMM4A and SMM4B will be discussed in detail in Section \ref{sec:discussion}. We present a summary of the results and our interpretation in Section \ref{sec:conclusions}. In addition, channel maps of the three molecular lines are shown in Appendix \ref{sec:app_ch}.

\section{ALMA OBSERVATIONS} \label{sec:observations}
We observed five regions in Serpens Main cluster, which was chosen from SMA archival data of a mosaicking survey carried out in 2010, using ALMA in its Cycle 3 on 2016 May 19 and 21. The results of the SMM11 condensation were reported in \citet{aso2017b}, and those of the SMM4 condensation are reported in this paper, while those of the other regions will be reported in future papers. 
Observing time for SMM4 including overhead is $\sim22$ min and $\sim 61$ min, while on-source observing time for SMM4 is 4.5 and 10.5 min in the first and the second days, respectively. The numbers of antenna were 37 and 39 in the first and the second days, respectively, and the antenna configuration of the second day was more extended than that of the first day. The minimum projected baseline length is 15 m. Any emission beyond $8\farcs0$ (3400 au) was resolved out by $\gtrsim 50\%$ with the antenna configuration \citep{wi.we1994}. Spectral windows for $^{12}$CO $(J=2-1)$, C$^{18}$O ($J=2-1$), and SO ($J_N=6_5-5_4$) lines have 3840, 1920, and 960 channels covering 117, 59, and 59 MHz band widths at frequency resolutions of 30.5, 30.5, and 61.0 kHz, respectively. In making maps, 32, 2, and 4 channels are binned for the $^{12}$CO, C$^{18}$O, and SO lines and the resultant velocity resolutions are 1.27, 0.083, and $0.33\ \kms$, respectively. Two other spectral windows cover 216-218 GHz and 232-234 GHz, which were assigned to the continuum emission.

All the imaging procedure was carried out with Common Astronomical Software Applications (CASA). The visibilities were Fourier transformed and CLEANed with Briggs weighting, a robust parameter of 0.0, and a threshold of 3$\sigma$. Multi-scale CLEAN was used to converge CLEAN, where CLEAN components were point sources or $\sim 1\farcs5$ Gaussian sources.

We also performed self-calibration for the continuum data using tasks in CASA ($clean$, $gaincal$, and $applycal$).
Only the phase was calibrated first with the time bin of 3 scans ($\sim 18$s). Then, using the derived table, the amplitude and the phase were calibrated together. The self-calibration improved the rms noise level of the continuum maps by a factor of $\sim 2$. The obtained calibration tables for the continuum data were also applied to the line data. The noise level of the line maps were measured in emission-free channels. The parameters of our observations mentioned above and others are summarized in Table \ref{ch4:tab:obs}.

\begin{deluxetable*}{ccccc}
\tablecaption{Summary of the ALMA observational parameters \label{ch4:tab:obs}}
\tablehead{
\colhead{Date} & \multicolumn{4}{c}{2016.May.19, 21}\\
\colhead{Projected baseline length} & \multicolumn{4}{c}{15 - 613 m (11 - 460 k$\lambda$)}\\
\colhead{Primary beam} & \multicolumn{4}{c}{$27\arcsec$}\\
\colhead{Passband calibrator} & \multicolumn{4}{c}{J1751$+$0939}\\
\colhead{Flux calibrator} & \multicolumn{4}{c}{Titan}\\
\colhead{Gain calibrator} & \multicolumn{3}{c}{J1830$+$0619 (470 mJy), J1824$+$0119 (79 mJy)}\\
\colhead{Coordinate center (J2000)} & \multicolumn{4}{c}{$18^{\rm h}30^{\rm m}$00\fs 38, $1^{\circ}11\arcmin 44\farcs 55$}\\
}
\startdata
 & Continuum & $^{12}$CO ($J=2-1$) & C$^{18}$O ($J=2-1)$ & SO ($J_N=6_5-5_4$)\\
\hline
Frequency (GHz) & 225 & 230.538000 & 219.560358 & 219.949433\\
Bandwidth/velocity resolution & 4 GHz & $1.27\ \kms$ & $0.083\ \kms$ & $0.33\ \kms$\\
Beam (P.A.) & $0\farcs 57\times 0\farcs46\ (-85\arcdeg)$ & $0\farcs 61\times 0\farcs 50\ (-82\arcdeg)$ & $0\farcs 64\times 0\farcs 52\ (-83\arcdeg)$ & $0\farcs 65\times 0\farcs 52\ (-85\arcdeg)$\\
rms noise level ($\mJB$) & 0.1 & 3.7 & 12 & 8.0\\
\enddata
\end{deluxetable*}


\section{RESULTS} \label{sec:results}
\subsection{1.3 mm continuum} \label{sec:cont} 
\begin{figure}[ht!]
\epsscale{1}
\plotone{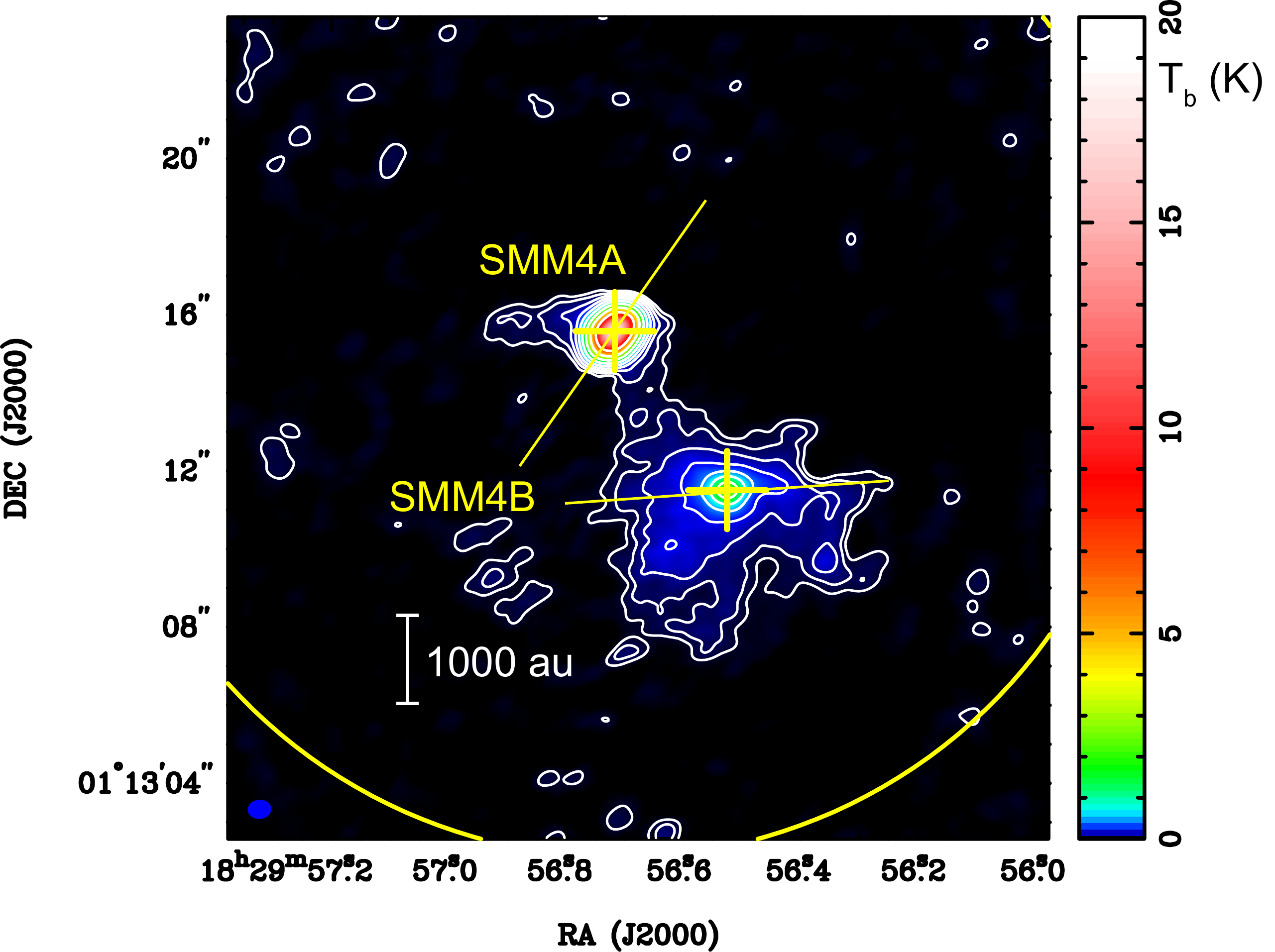}
\caption{Continuum emission map of SMM4A and SMM4B. Contour levels are $3,6,12,24,\dots \times \sigma$, where 1$\sigma$ corresponds to $0.1\ \mJB$. The color map shows brightness temperature in the unit of K. Yellow plus signs and lines denote two peak positions and major axes derived from 2D Gaussian fitting. Yellow circle denotes ALMA primary beam (FWHM$\sim 27\arcsec$). A blue-filled ellipse at the bottom-left corner denotes the ALMA synthesized beam; $0\farcs 57\times 0\farcs 46,\ {\rm P.A.}=-85^{\circ}$. 
\label{fig:cont}}
\end{figure}

Figure \ref{fig:cont} shows a map of the 1.3 mm continuum emission in the SMM4 condensation. The emission has two local peaks. We label the stronger and the weaker peaks as SMM4A and SMM4B, respectively, in this paper. SMM4A shows a size of $\sim 2\arcsec \ (860$ AU) at the $3\sigma$ noise level, while SMM4B shows a $\sim 5\arcsec \ (2100$ AU) size at the same noise level. The peak positions of SMM4A and SMM4B measured by Gaussian fitting were $\alpha ({\rm J2000})=18^{\rm h}29^{\rm m}56\fs 718,\ \delta ({\rm J2000})=1\arcdeg 13\arcmin 15\farcs 58$ and $\alpha ({\rm J2000})=18^{\rm h}29^{\rm m}56\fs 526, \delta ({\rm J2000})=1\arcdeg 13\arcmin 11\farcs 50$, respectively. The fitting uses emission within the $3\sigma$ contour enclosing each peak after primary beam correction, providing peak intensities and deconvolved sizes (FWHM), as well as the positions. Total flux densities of the sources were measured by integrating the flux in the regions enclosed by the $3\sigma$ contours after primary beam correction.

The emission around SMM4A shows a compact structure with extensions to the east and the south. Its peak intensity, total flux density, and deconvolved size are $205\ \mJB$, $492\ \mJ$, and $0\farcs 75 \times 0\farcs 46$ (P.A.=145$\arcdeg$), respectively. The peak intensity corresponds to a brightness temperature of $T_{\rm b}=18$ K. The dust temperature in Serpens Main was estimated by \citet{leKI2014} to be $\sim 20$ K. Thus, the continuum emission in SMM4A is thought to be optically thick. A lower limit to the gas mass can be calculated from the total flux density to be $0.83\ \Ms$ by assuming optical thinness, a dust temperature $T_{\rm dust}=20$ K, opacity coefficient $\kappa(850\ \micron)=0.035\ {\rm cm}^{2}~{\rm g}^{-1}$ \citep{an.wi2005}, opacity index $\beta=1$, and a gas-to-dust mass ratio of 100.

In contrast to SMM4A, the emission from SMM4B shows a more extended structure at lower contour levels, as well as a compact structure at higher contour levels. The extended structure mainly consists of components toward the southeast and west from the SMM4B position. The peak intensity, total flux density, and deconvolved size of the SMM4B continuum emission are $25\ \mJB$, $173\ \mJ$, and $0\farcs 70 \times 0\farcs 53$ (P.A.=95$\arcdeg$), respectively. The peak intensity corresponds to a brightness temperature of $T_{\rm b}=2.0$ K. The total flux density corresponds to a gas mass of $0.29\ \Ms$, while the peak intensity corresponds to a gas mass column density of $0.04\ \Ms~{\rm beam}^{-1}$ under the same assumptions as above.

\subsection{Spectral Energy Distribution} \label{sec:inf}
\begin{figure}[ht!]
\epsscale{1.2}
\plotone{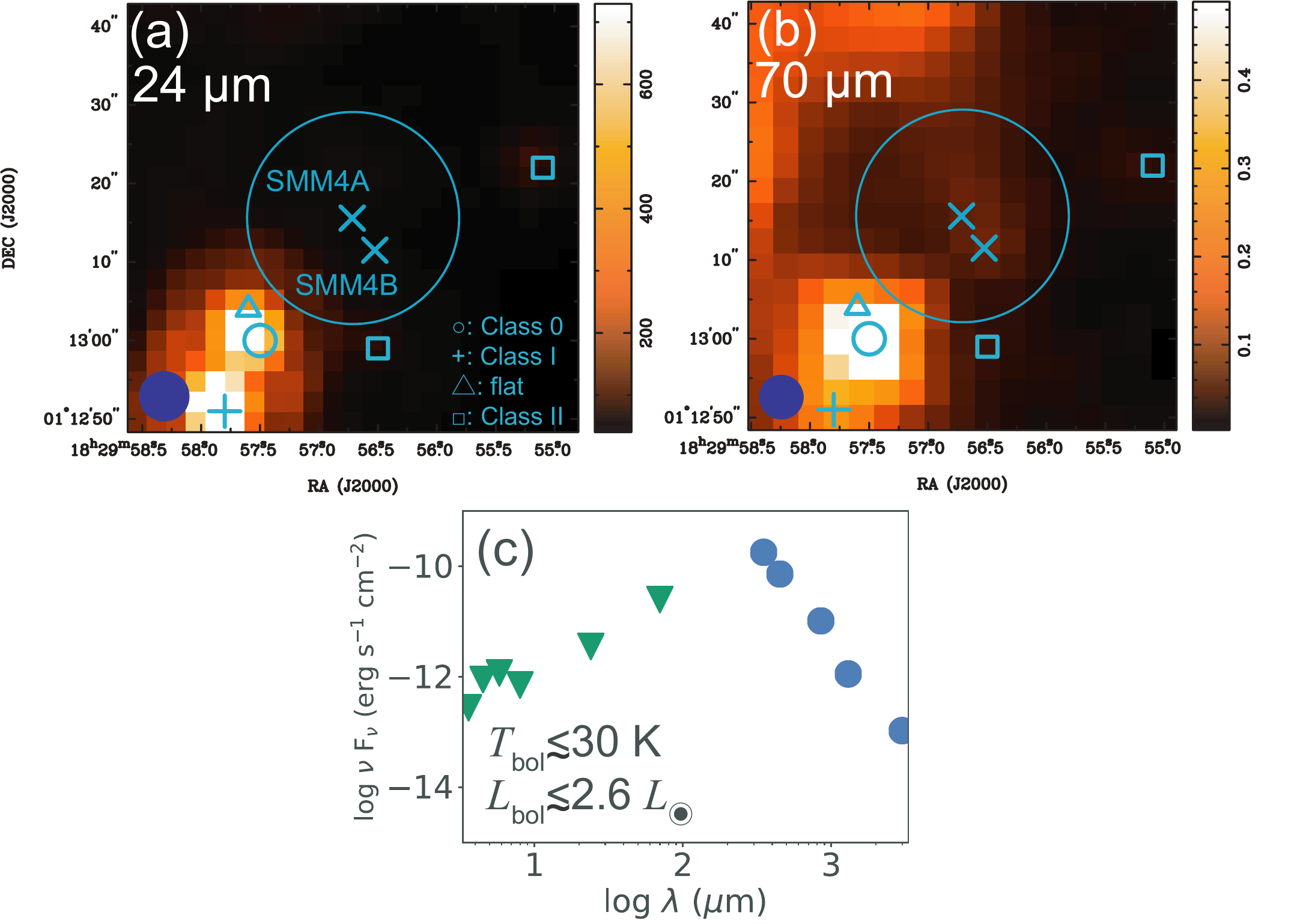}
\caption{Infrared observations toward SMM4A and SMM4B. (a) $Spitzer$ 24 $\micron$ image in ${\rm Jy}~{\rm sr}^{-1}$. (b) $Herschel$ 70 $\micron$ image in ${\rm Jy}~{\rm pixel}^{-1}$, where 1 pixel is $3\farcs2 \times 3\farcs2$. Blue filled ellipses at each bottom-left corner denote the point spread functions: $6\farcs4$ and $5\farcs7$ for $Spitzer$ 24 $\micron$ and $Herschel$ 70 $\micron$, respectively. X marks denote the peak positions in 1.3 mm. Green marks denote positions of YSOs identified by $Spitzer$ observations \citep{du2015}. Large circles denote the ALMA primary beam (FWHM$\sim27\arcsec$). (c) SED of SMM4. Blue points denote measured flux density, while green points denote our detection limits.
\label{fig:infra}}
\end{figure}

Figure \ref{fig:infra}a and \ref{fig:infra}b show {\it Spitzer} $24\,\micron$ and {\it Herschel} $70\,\micron$ images of the SMM4 region. SMM4A and SMM4B are not detected by {\it Spitzer} and only marginally with {\it Herschel} where they are much fainter than a neighboring Class 0 protostar. Figure \ref{fig:infra}c shows the SED of the overall submillimeter condensation SMM4 as defined by the JCMT and derived from $Spizter$ IRAC (3.6, 4.5, 5.8, 8.0 $\micron$), $Spizter$ MIPS 24 $\micron$, $Herschel$ PACS 70 $\micron$ observations, ALMA 1.3 mm (this work), and the literature: CSO SHARK-$\II$ 350 $\micron$ \citep{su2016}, JCMT SCUBA 450 $\micron$ and 850 $\micron$ \citep{da1999}, and CARMA 3 mm \citep{leKI2014}. The upper limits in the SED, which are set to be the flux densities leaked from protostars near SMM4A and SMM4B, are shown with green marks in Figure \ref{fig:infra}. These detection limits are higher than three times the statistical noise levels. Because SMM4A and SMM4B are not spatially resolved in many of the mid-infrared to submillimeter observations mentioned above, flux densities were measured within apertures whose diameter is twice larger than the full width at half maximum (FWHM) of each point spread function (PSF) at wavelengths from 3.6 to $70\ \micron$, while peak flux densities were taken from the literature at the other wavelengths.

The bolometric temperature and luminosity of SMM4 were directly calculated to be $T_{\rm bol}\lesssim $30 K and $L_{\rm bol}\lesssim 2.6\ \Ls$ by the trapezoidal integration, rather than fitting, from this SED. We include the upper limits at 70 $\micron$ and shorter, so the derived values are upper limits. This bolometric temperature is in the Class 0 range \citep[$T_{\rm bol}<70$ K;][]{ch1995}, but the individual temperatures of SMM4A and SMM4B may be different because they are not spatially resolved. Bolometric temperatures can also be underestimated when YSOs have edge-on disks with significant sizes \citep{jo2009}. Thus other quantities are required to constrain their evolutionary phase. A submillimter luminosity is similarly calculated to be $L_{\rm submm}\sim 0.3\ \Ls$ by integrating fluxes from 350 $\micron$ to 3 mm. $L_{\rm bol}/L_{\rm submm}\lesssim 9$ is also lower than a threshold, $L_{\rm bol}/L_{\rm submm}\sim 200$ \citep{an1993}, between Class 0 and Class I, and even lower than that of typical Class 0 protostars \citep{gr2013,aso2015}. In addition, an internal luminosity is estimated from the 70 $\micron$ flux density to be $L_{\rm int}\lesssim 0.3\ \Ls$ using the empirical relation reported by \citet{du2008}. For these reasons, we consider SMM4A and SMM4B to be protostars formed in the same core SMM4 identified with JCMT observations, similar to the VLA1623 pair \citep{mu.la2013}. 

\subsection{$^{12}$CO and SO lines} \label{sec:12coso}
\begin{figure*}[ht!]
\gridline{
\fig{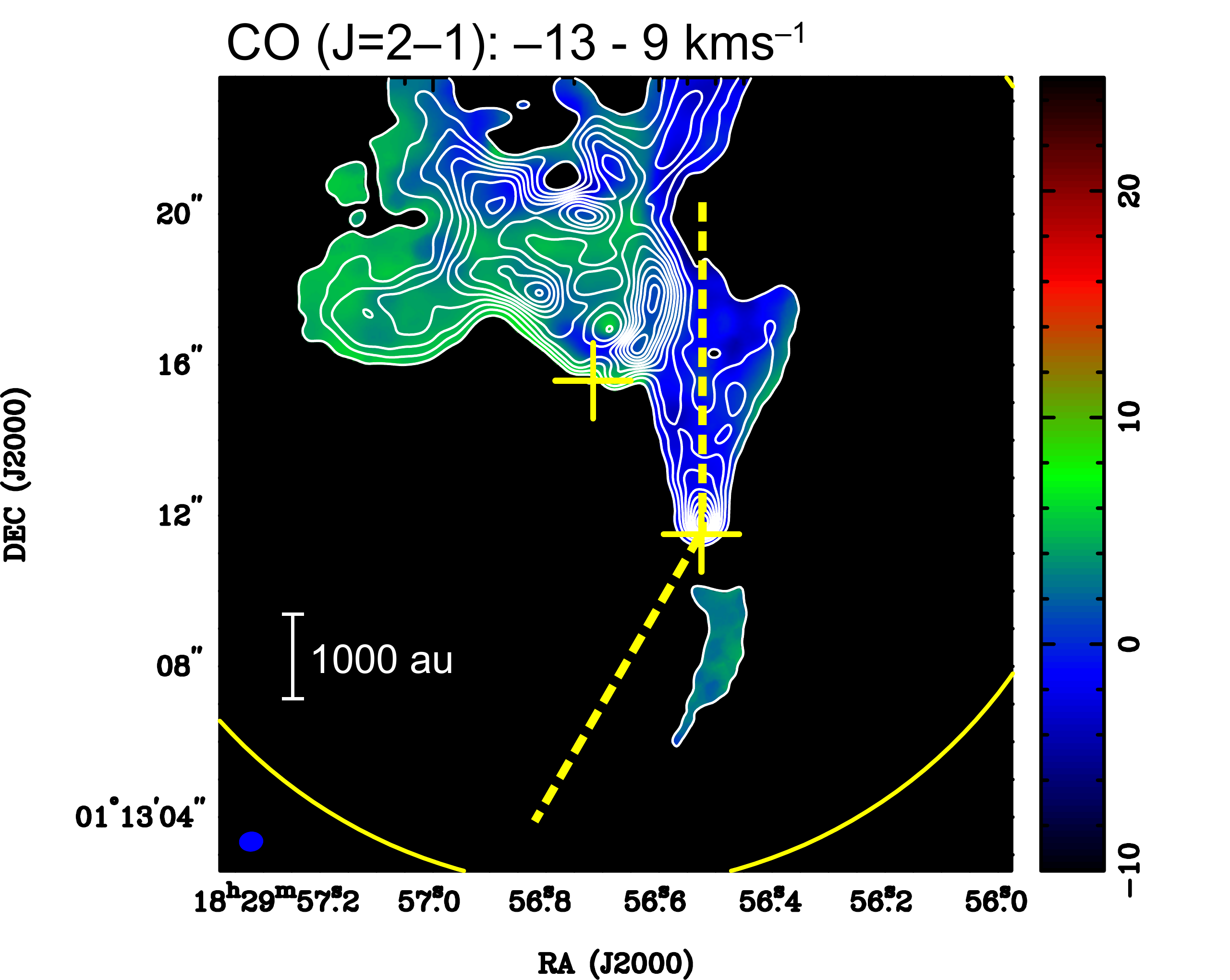}{0.4\textwidth}{(a)}
\fig{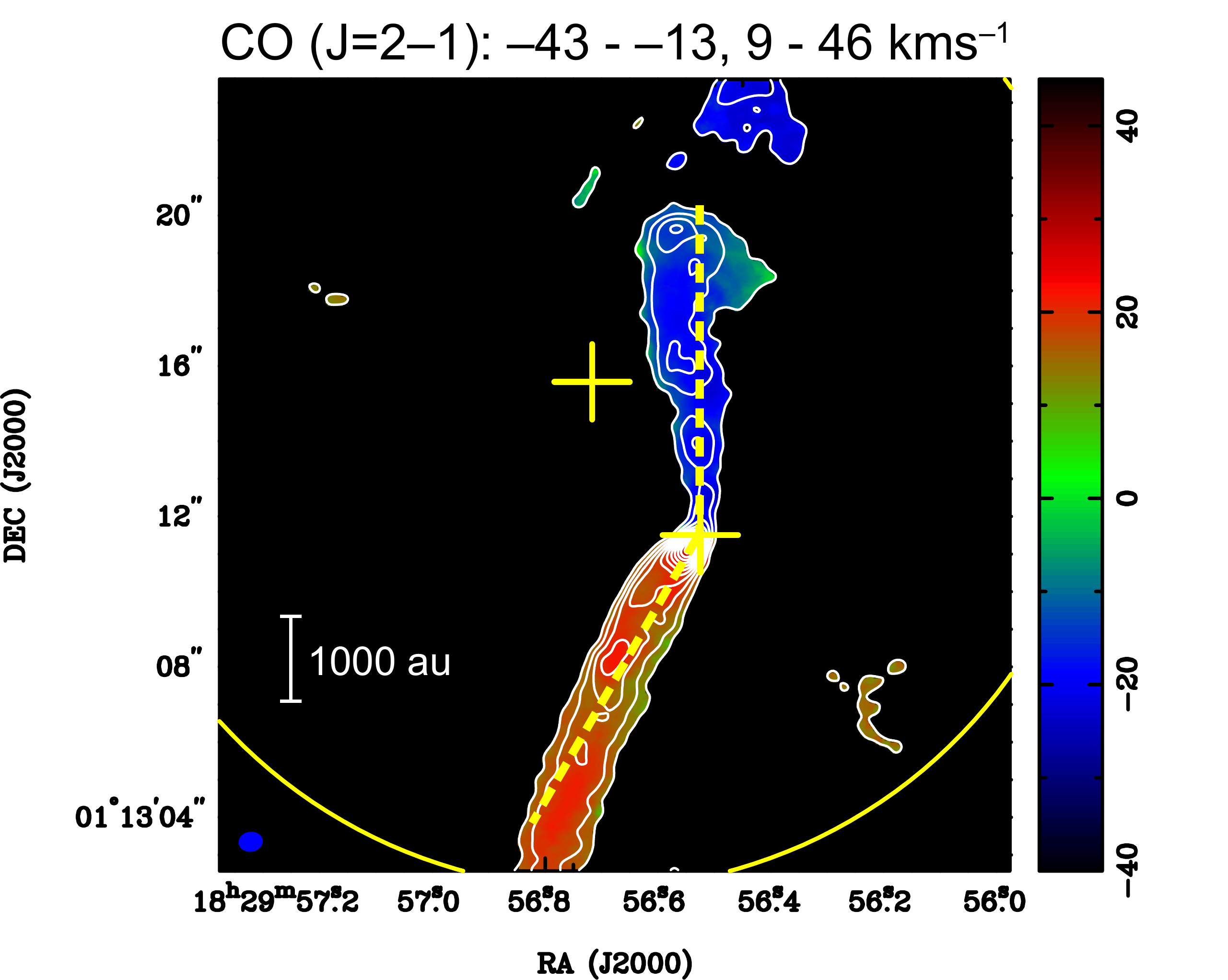}{0.4\textwidth}{(b)}
}
\gridline{
\fig{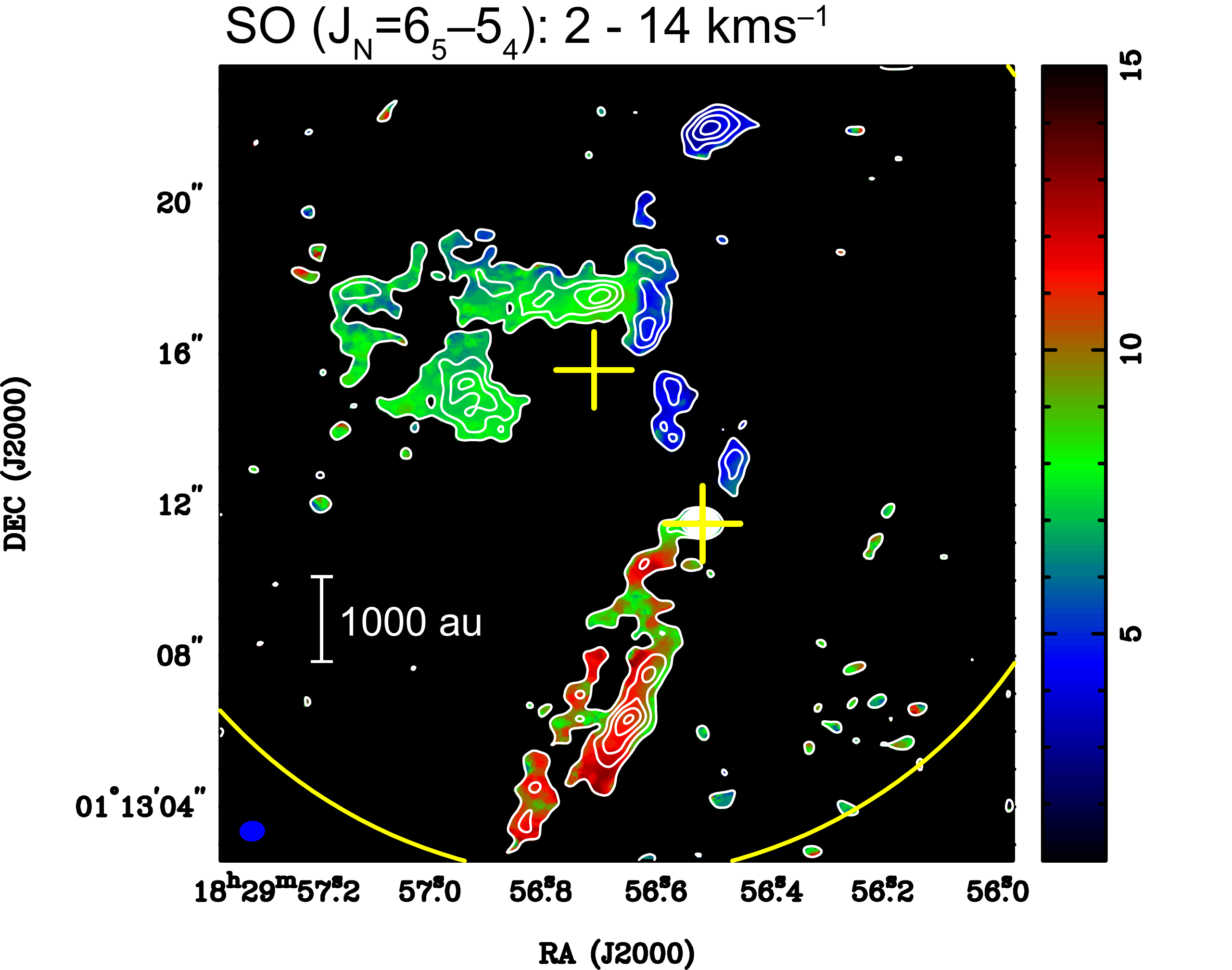}{0.4\textwidth}{(c)}
\fig{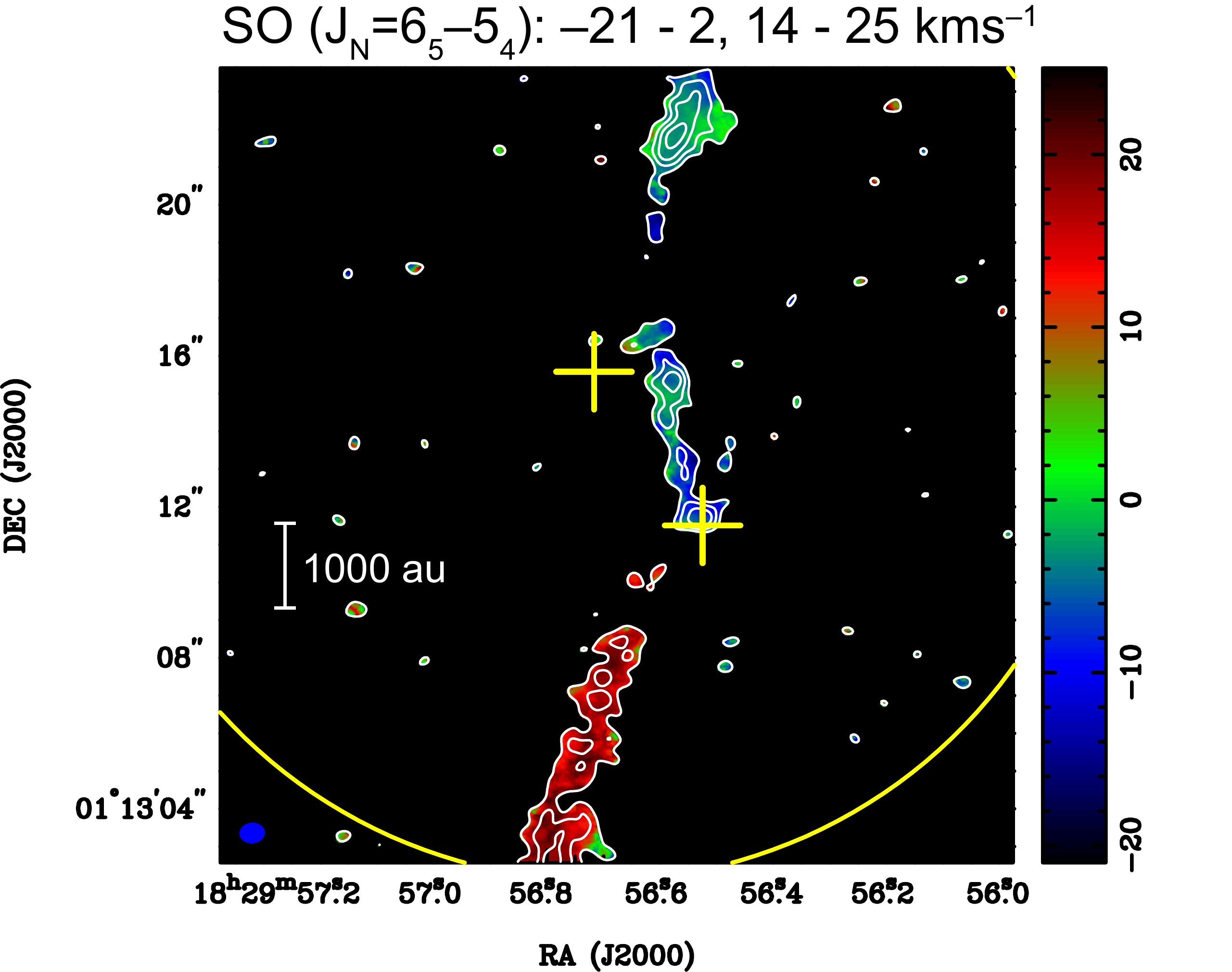}{0.4\textwidth}{(d)}
}
\caption{Integrated intensity (moment 0; contours) and mean velocity (moment 1; color) maps in (a) $^{12}$CO lower, (b) $^{12}$CO higher, (c) SO lower, and (d) SO higher velocity components of SMM4A and SMM4B. The integrated velocity range of each component is written in the upper part of each panel. Contour levels are: (a) from $20\sigma$ in steps of $20\sigma$, (b) from $10\sigma$ in steps of $10\sigma$, (c) from $3\sigma$ in steps of $3\sigma$, and (d) from $3\sigma$ in steps of $3\sigma$, where $1\sigma$ corresponds to (a) 20, (b) 35, (c) 16, and (d) $27\ \mJB~\kms$. Yellow plus signs and circle are the same as in Figure \ref{fig:cont}. Blue-filled ellipses at the bottom-left corners denote the ALMA synthesized beams: $0\farcs 61\times 0\farcs 50,\ {\rm P.A.}=-82^{\circ}$ for $^{12}$CO emission and $0\farcs 65\times 0\farcs 52,\ {\rm P.A.}=-85^{\circ}$ for SO emission. Yellow dashed lines denote directions of P.A.=0$\arcdeg$ and 150$\arcdeg$, which is used to make a position-velocity diagram (Figure \ref{fig:curved}b).
\label{fig:12coso}}
\end{figure*}

Figure \ref{fig:12coso}a and \ref{fig:12coso}b show moment 0 and 1 maps of low- and high-velocity components of the $^{12}$CO emission line, respectively, around SMM4A and SMM4B. Figure \ref{fig:12coso}c and \ref{fig:12coso}d show those of the SO emission line. The integrated velocity range of the $^{12}$CO low-velocity component is decided to highlight $^{12}$CO emission associated with SMM4A, while that of the SO low-velocity component is chosen to highlight SO emission associated with SMM4A and shell structures associated with SMM4B. The SO emission associated with SMM4A is detected from $V_{\rm LSR}=6$ to $9\ \kms$. More detailed velocity structures can be seen in channel maps of the $^{12}$CO emission (Figure \ref{fig:coch}) and the SO emission (Figure \ref{fig:soch}) shown in Appendix \ref{sec:app_ch}.

The $^{12}$CO emission overall traces outflows associated with SMM4A and SMM4B. Figure \ref{fig:12coso}a shows that SMM4A has a blueshifted unipolar outflow. The absence of its redshifted counterpart in the south may be due to less material on the southern side and/or resolving-out of the emission from the counterpart. We note that CARMA observations with a $\sim 8\arcsec$ resolution show less dense gas in N$_2$H$^+$ ($1-0$) in the south of SMM4 than in the other directions \citep{leKI2014}, while an extended redshifted component is detected in the south of SMM4 by JCMT in CO $J=2-1$ \citep{da1999}. A blueshifted lobe associated with SMM4B is also detected in the low velocity range. The blueshifted lobe from SMM4A shows a fan-shaped structure, while that from SMM4B shows a relatively collimated structure. In addition, the integrated intensity is enhanced on the northwestern side of SMM4A, where the SMM4A outflow appears to overlap with the SMM4B blueshifted lobe. Figure \ref{fig:12coso}b shows that the high-velocity component traces only the SMM4B outflow. In this velocity range, both blue- and redshifted lobes are collimated with a length of $\sim 10\arcsec$ (4300 au) and a width of $\sim 1\arcsec$ (430 au). The SMM4B outflow shows knots at high velocities, $V_{\rm LSR}\lesssim -35\ \kms,\ 15\ \kms \lesssim V_{\rm LSR}$ (Figure \ref{fig:coch}). Velocity structures of the SMM4B $^{12}$CO outflow will be investigated in more detail in Section \ref{sec:conf}.

Figure \ref{fig:12coso}c shows that the SO low-velocity component traces the blue- and redshifted lobes of the SMM4B outflow and another component surrounding SMM4A. Compared to the $^{12}$CO emission tracing the SMM4B outflow, the SO emission is enhanced at the cavity walls of the two lobes, including a strongly emitting blob $\sim 10\arcsec$ (4100 au) away from the SMM4B in the north. Figure \ref{fig:12coso}d shows that the high-velocity component traces the eastern edges of the lobes of the SMM4B outflow. In addition, the SO emission is also enhanced in the vicinity of SMM4B in both low- and high-velocity components. The SO enhancement at the cavity walls can be interpreted as shocked regions where the outflow is interacting with an envelope around SMM4B. The SO enhancement at the SMM4B position, meanwhile, could be due to accretion shock, as will be discussed in Section \ref{sec:evo}.

\subsection{C$^{18}$O line} \label{sec:c18o}
\subsubsection{Overall Structures and Systemic Velocities} \label{sec:c18ooverall}
\begin{figure}[ht!]
\epsscale{1}
\plotone{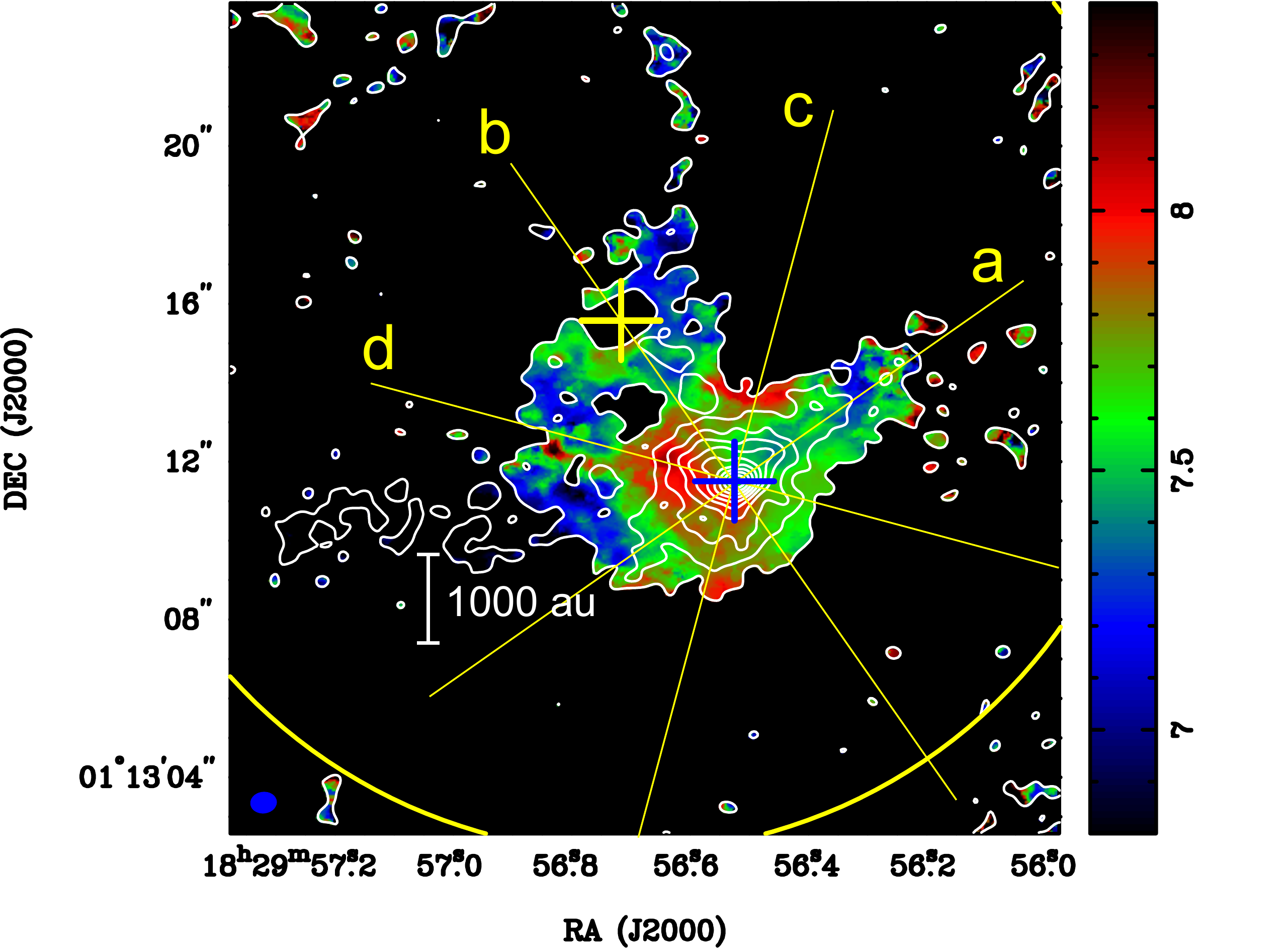}
\caption{Integrated intensity (moment 0; contours) and mean velocity (moment 1; color) maps in the C$^{18}$O emission of SMM4A and SMM4B. The integrated velocity range is from $6$ to $9\ \kms$. Contour levels of the integrated intensity maps are from $3\sigma$ in steps of $5\sigma$, where $1\sigma$ corresponds to $6\ \mJB~\kms$. Plus signs and circle are the same as in Figure \ref{fig:cont}. A blue-filled ellipse at the bottom-left corner denotes the ALMA synthesized beam; $0\farcs 64\times 0\farcs 52,\ {\rm P.A.}=-83^{\circ}$. Yellow solid lines passing SMM4B show cuts for position-velocity (PV) diagrams (Figure \ref{fig:c18opv}).
\label{fig:c18o}}
\end{figure}

Figure \ref{fig:c18o} shows moment 0 and 1 maps of the C$^{18}$O emission. Strong emission is detected at the SMM4B position, while negative intensities are observed at the SMM4A position, surrounded by weak positive intensities. Although Figure \ref{fig:c18o} does not show a negative contour, it can be seen in channel maps (Figure \ref{fig:c18och}) in Appendix \ref{sec:app_ch}. The negative intensities are due to a continuum subtraction effect, which can occur when line emission traces gas with a lower temperature than the brightness temperature of background continuum emission at the Rayleigh-Jeans limit. In fact, the brightness temperature of the dust continuum emission is higher in SMM4A than in SMM4B as seen in Figure \ref{fig:cont}. The C$^{18}$O emission around SMM4B is extended on a $\sim 4\arcsec$ (1700 au) scale with an extension to the northwest. A velocity gradient can be seen from the southeast to the northwest of SMM4B, which is not the same as one seen in the $^{12}$CO emission.

\begin{figure}[ht!]
\gridline{
\hspace{-2em}
\fig{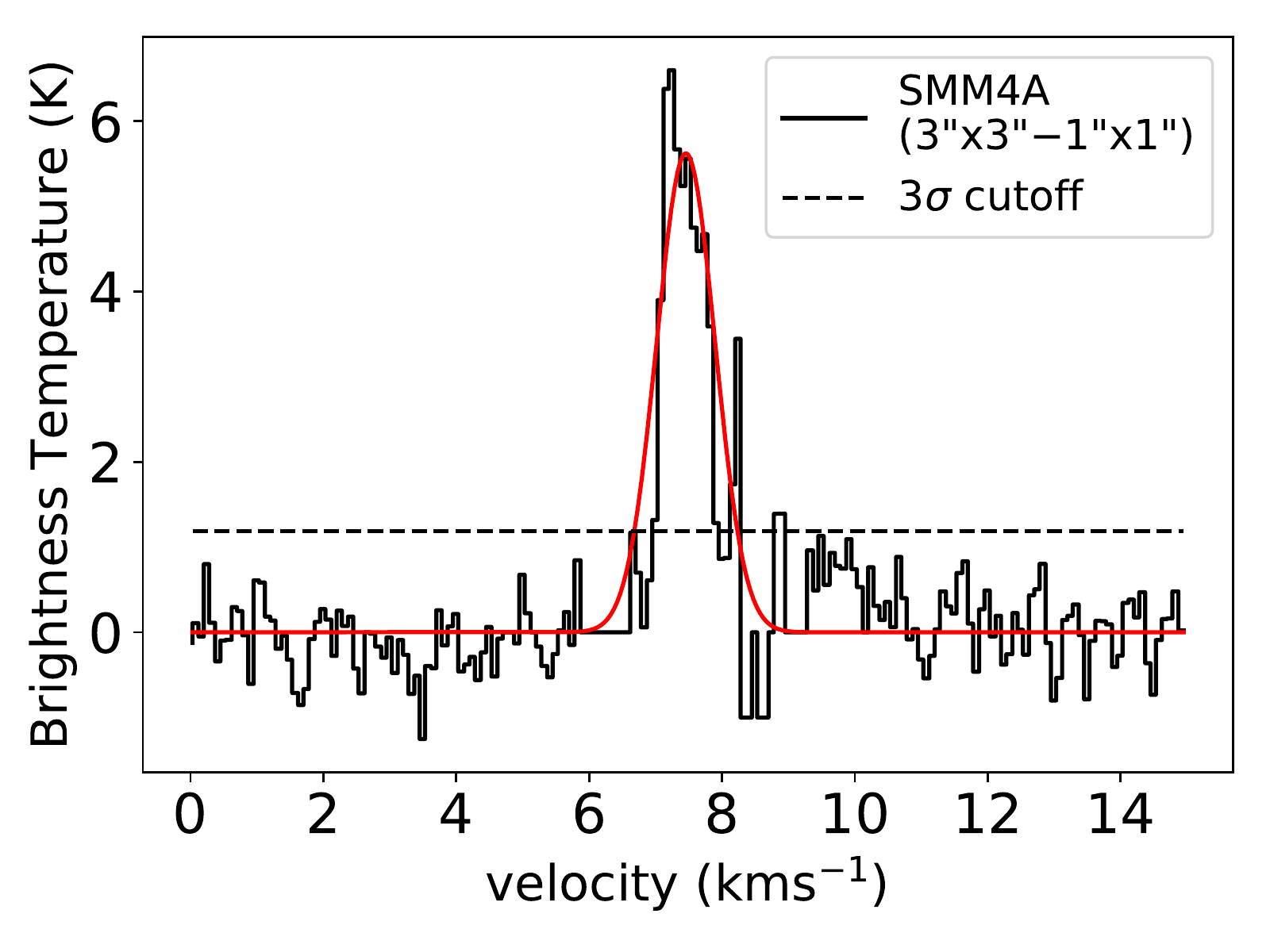}{0.25\textwidth}{(a)}
\fig{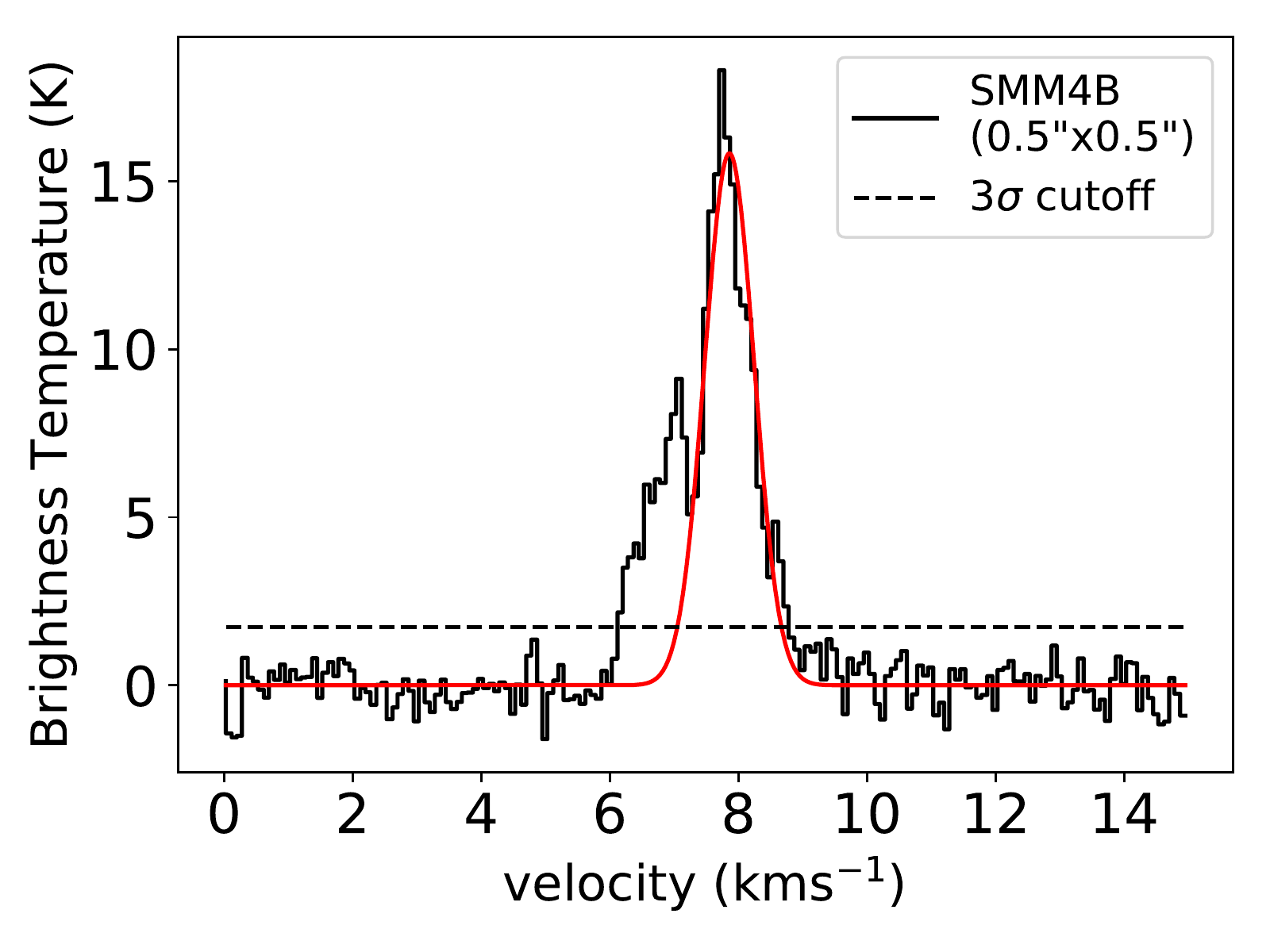}{0.25\textwidth}{(b)}
}
\caption{
C$^{18}$O line profiles of (a) SMM4A and (b) SMM4B. Horizontal dashed lines and red curves denote 3$\sigma$ cutoffs and the best-fit Gaussian profiles, respectively, in each panel. (a) The spectrum was made within the $3\arcsec \times 3\arcsec $ region centered at SMM4A except for the $1\arcsec \times 1\arcsec$ region centered at SMM4A, where the C$^{18}$O line has negative intensities. The best-fit Gaussian profile is described by a peak temperature of 5.6 K, a mean velocity of $7.46\ \kms$, and a FWHM velocity width of $1.0\ \kms$. (b) The spectrum was made within the $0\farcs 5 \times 0\farcs 5$ region centered at SMM4B. The main component at $V>7.35\ \kms$ was used for the fitting. The best-fit Gaussian profile is described by a peak temperature of 15.8 K, a mean velocity of $7.86\ \kms$, and a FWHM velocity width of $0.91\ \kms$.
\label{fig:c18opro}}
\end{figure}

To investigate the velocity structures of the C$^{18}$O emission, two other types of figures are inspected. First we determine the systemic velocities of SMM4A and SMM4B using line profiles of the C$^{18}$O emission shown in Figure \ref{fig:c18opro}. The line profile of SMM4A (Figure \ref{fig:c18opro}a) was derived from the central $3\arcsec \times 3\arcsec$ region excluding the central $1\arcsec \times 1\arcsec$ in SMM4A in order to avoid the negative intensities. The line profile of SMM4B (Figure \ref{fig:c18opro}b) was, meanwhile, derived from the central $0\farcs 5\times 0\farcs 5$ region in SMM4B. The SMM4B profile (Figure \ref{fig:c18opro}b) shows two components along the velocity direction. We thus only use the component in $V>7.35\ \kms$ to determine the systemic velocity of SMM4B; it will be revealed below that this component represents the C$^{18}$O emission in SMM4B.

The two profiles were fitted with a Gaussian function using emission above $3\sigma$ levels denoted with horizontal dashed lines in Figure \ref{fig:c18opro}. The fitting to the SMM4A profile provides a mean velocity of $7.46\pm 0.02\ \kms$. This velocity is close to that of a negative-intensity component in SMM4A (see Section \ref{sec:c18opv}). Other parameters derived by the Gaussian fitting to SMM4A are a peak brightness temperature of $5.6\pm0.6$ K and an FWHM velocity width of $1.0\pm 0.2\ \kms$. Similarly the Gaussian mean velocity of SMM4B is $7.86\pm 0.02\ \kms$, its peak brightness temperature is $15.8\pm 0.7$ K, and its FWHM velocity width is $0.91\pm 0.06\ \kms$. We adopt the Gaussian mean velocities as the systemic velocities of SMM4A and SMM4B.

\subsubsection{Position-Velocity Diagrams and Velocity Gradients} \label{sec:c18opv}
\begin{figure*}[ht!]
\gridline{
\fig{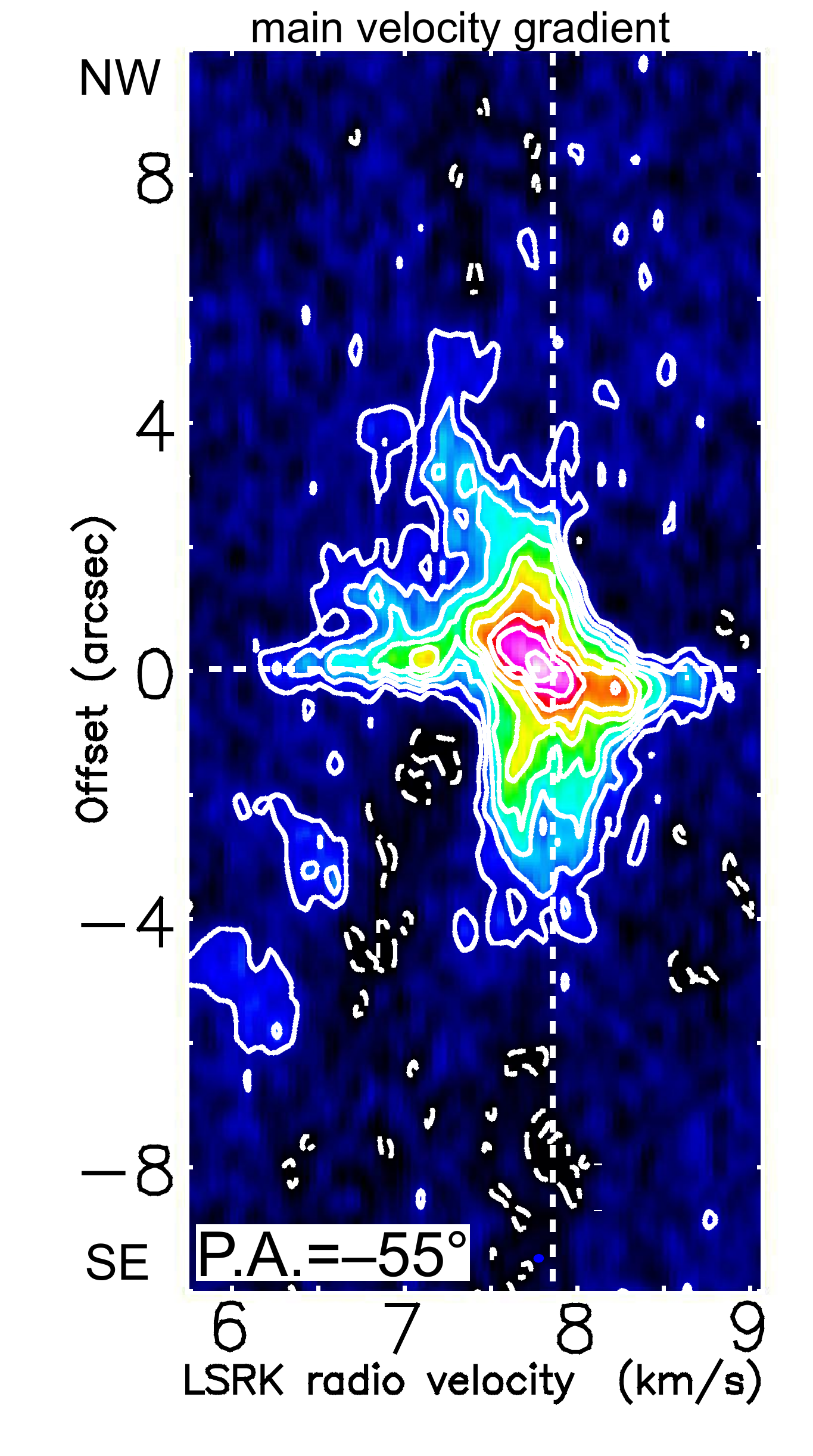}{0.24\textwidth}{(a)}
\fig{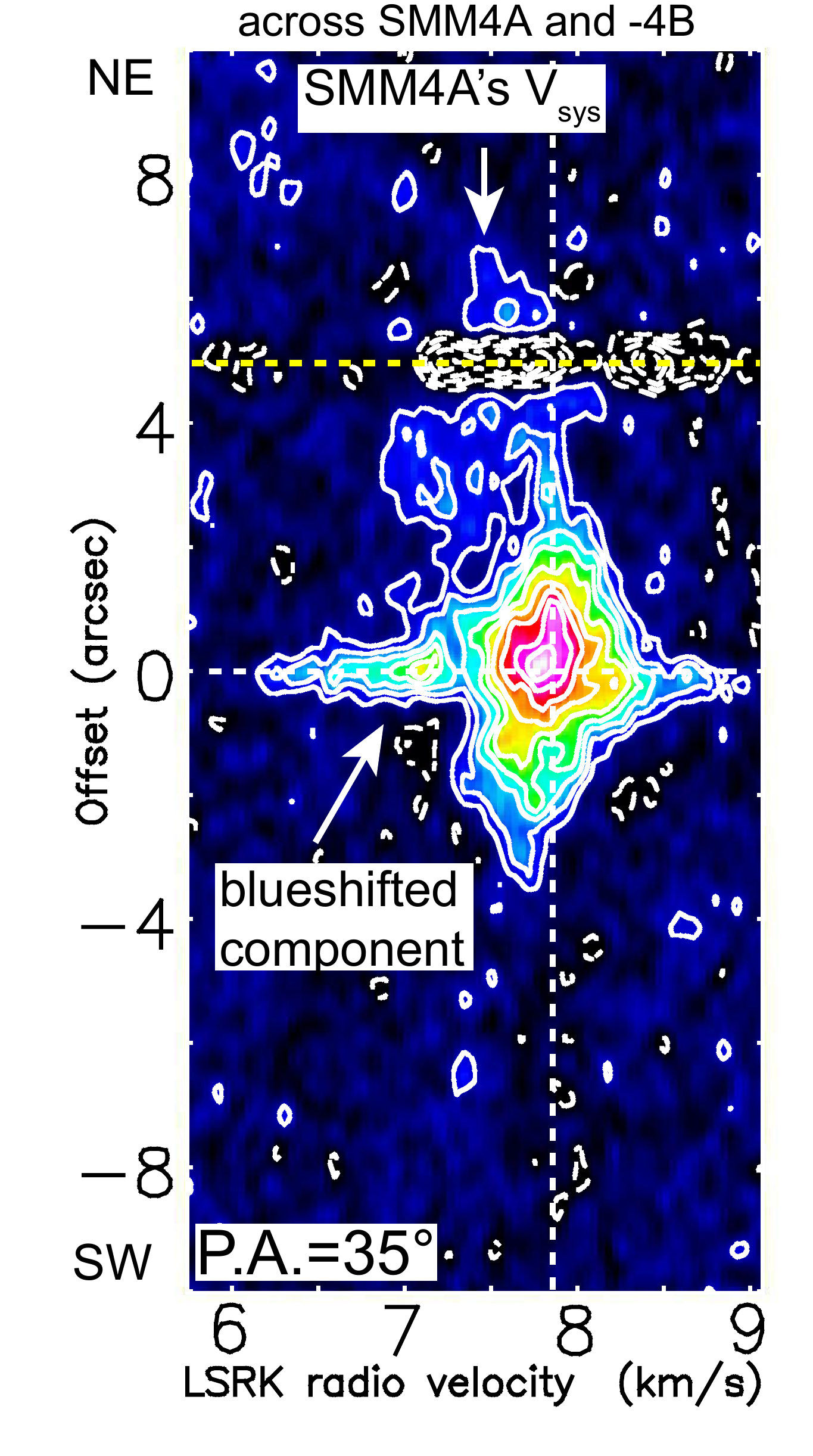}{0.24\textwidth}{(b)}
\fig{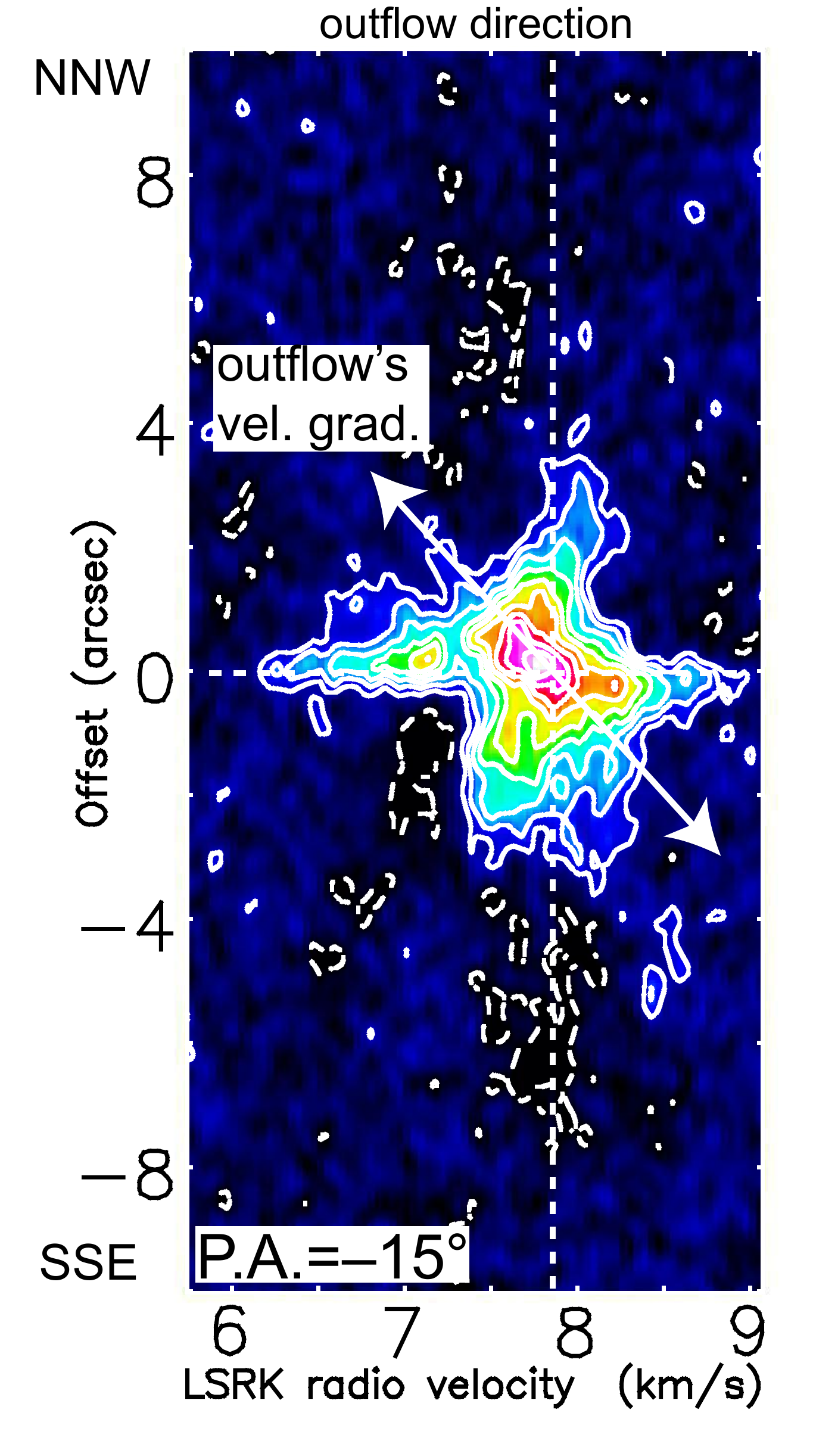}{0.24\textwidth}{(c)}
\fig{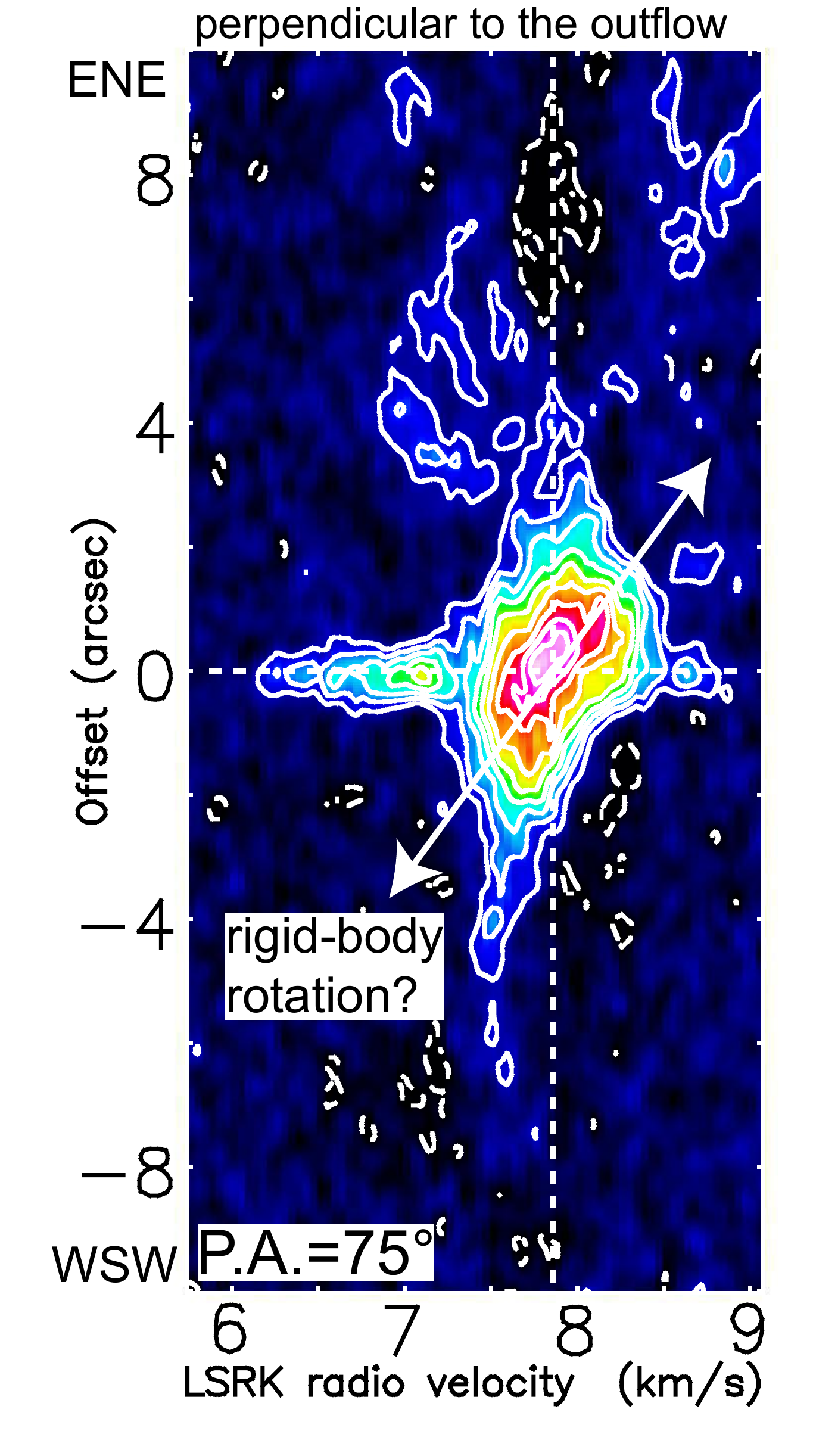}{0.24\textwidth}{(d)}
}
\caption{
Position-velocity diagrams of the C$^{18}$O emission in SMM4B along (a) the main velocity gradient (P.A.=$-55\arcdeg$), (b) the line across SMM4A and SMM4B (P.A.=35$\arcdeg$), (c) the central axis of the two outflow lobes (P.A.=$-15\arcdeg$), and (d) a perpendicular direction of the outflow (P.A.=75$\arcdeg$). The width of cut is $0\farcs 55$ (1 beam). Contour levels are from $3\sigma$ in steps of $3\sigma$, where $1\sigma$ corresponds to $8.5\ \mJB$. Positive offsets correspond to the northern side of SMM4B. White vertical and horizontal dashed lines denote the systemic velocity ($V_{\rm LSR}=7.86\ \kms$) and the SMM4B position, respectively, while the yellow horizontal dashed line in panel (b) denotes the SMM4A position. 
\label{fig:c18opv}}
\end{figure*}

Next we investigate the velocity gradients around SMM4B using PV diagrams. Figure \ref{fig:c18opv} shows PV diagrams along (a) the direction perpendicular to the line passing through SMM4A and SMM4B, P.A.=$-55\arcdeg$, (b) the direction passing through SMM4A and SMM4B, P.A.=35$\arcdeg$, (c) the central axis of the blue- and redshifted lobes of the SMM4B outflow, P.A.=$-15\arcdeg$, and (d) the direction perpendicular to the outflow direction, P.A.=$75\arcdeg$. The direction (a) is also along the main velocity gradient seen in the mean velocity map (Figure \ref{fig:c18o}). In addition to this gradient around the systemic velocity, Figure \ref{fig:c18opv}a also shows a compact component at $V_{\rm LSR}=7\ \kms$. These two components are also seen in all of the other PV diagrams (Figure \ref{fig:c18opv}b-d).

Figure \ref{fig:c18opv}b shows negative intensities at the SMM4A position (offset$=5\arcsec$). The negative intensities consist of two components along the velocity direction; the component at $7\ \kms \lesssim V\lesssim 8\ \kms$ appears to be associated to emission surrounding SMM4A, while the more redshifted component ($8\ \kms\lesssim V\lesssim 9\ \kms$) has a different velocities than the SMM4A main emission. The negative peak intensities of the two components correspond to $\sim-9$ K with respect to the continuum level. Figure \ref{fig:c18opv}b also shows two components associated with SMM4B; a compact component at velocities $<7.35\ \kms$ with a size of $\sim 0.5\arcsec$ (210 au) and a more extended brighter component at velocities $>7.35\ \kms$ with a size of $\sim 6\arcsec$ (2600 au), decreasing at higher velocities. A small velocity gradient from a blueshifted-southern part to a redshifted-northern part can be seen in this PV diagram though it is not as pronounced as that in the other PV diagrams (Figure \ref{fig:c18opv}a, \ref{fig:c18opv}c, and \ref{fig:c18opv}d).

The main velocity gradient seen in the C$^{18}$O line is neither parallel nor perpendicular to the associated $^{12}$CO outflow. The bending outflow also complicates the geometry of this system and the uncertain orientation and inclination angles prevent us from measuring kinematic quantities such as rotation and infall motion. We thus only mention two interpretations of the velocity gradient in the C$^{18}$O line. One is a rotating and infalling envelope whose rotational axis is parallel to the associated outflow. Figure \ref{fig:c18opv}c overall shows a velocity gradient along the outflow direction, axis (c) in Figure \ref{fig:c18o}, from the blueshifted components on the northern side to the redshifted components on the southern side, including compact components in the higher velocities. The gradient direction is similar to the one seen in the $^{12}$CO outflow; such a gradient can be observed if there is a disk-like infalling envelope. On the other hand, Figure \ref{fig:c18opv}c also shows that the northernmost part is redshifted and the southernmost part is blueshifted, which cannot be explained by the simple infalling motion. Figure \ref{fig:c18opv}d, a cut along the direction perpendicular to the outflow shows the blueshifted components on the southern side and the redshifted components on the northern side. Such a velocity gradient perpendicular to the outflow axis can be due to rotation of the envelope around SMM4B. If this is the case, the linear morphology of the C$^{18}$O emission passing the center of the PV diagram in Figure \ref{fig:c18opv}d appears more like rigid-body rotation than differential rotation on the scale of our angular resolution, $\sim 0\farcs55$ (240 au).
The other interpretation of the main velocity gradient seen in Figure \ref{fig:c18o} is the global rotation of the surrounding envelope, while its rotation axis is different from the outflow axis by $40\arcdeg-50\arcdeg$. This is possible because the outflow is launched on a scale smaller than circumstellar-disk scales, while the C$^{18}$O emission traces velocity gradients on the much larger envelope scale. Such misalignment of an envelope-rotation axis and an outflow axis is predicted in magneto-hydrodynamics (MHD) simulations in the context of misalignment between magnetic fields and initial rotation axes \citep[e.g.,][]{jo2012,li2013,ts2017,ma2017}.

\begin{figure}[ht!]
\epsscale{1}
\plotone{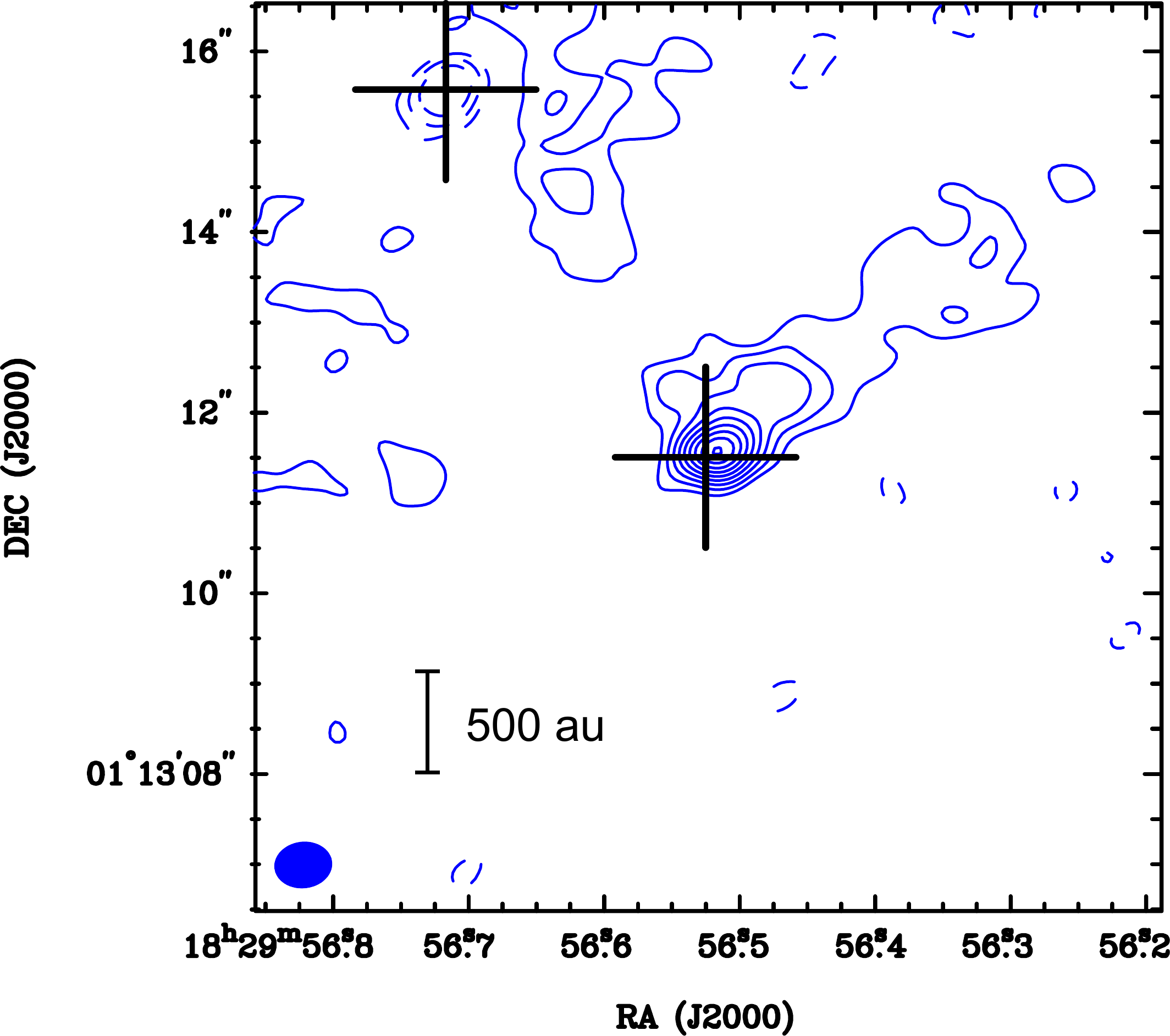}
\caption{Distribution of the blueshifted component of the C$^{18}$O emission in SMM4B integrated from $5.7$ to $7.35\ \kms$. Contour levels are from $5\sigma$ in steps of $3\sigma$, where $1\sigma$ corresponds to $4.4\ \mJB~\kms$. Plus signs and a filled ellipse at the bottom-left corner are the same as in Figure \ref{fig:cont}.
\label{fig:c18oblue}}
\end{figure}

The blueshifted ($V<7.35\ \kms$) component appears independent from the main component because it shows a local peak in the PV diagrams Figure \ref{fig:c18opv}a-d. Figure \ref{fig:c18oblue} shows the spatial distribution of this blueshifted component. It consists of a central compact component and extensions to the northwest and to the northeast. These northeastern and northwestern extensions are likely to be the walls of the V-shaped cavity of the SMM4B outflow. The central component is also slightly apart from the continuum peak position (plus sign) to the north, rather than the south. This direction is the same as that of blueshifted components of the SMM4B outflow. Hence, a possible interpretation of this blueshifted component is interaction between the envelope and the outflow. On the other hand, the highest velocity of the blueshifted component $\Delta V\sim 1.5\ \kms$ appears at very close $(r\sim 0\farcs 5)$ to the central stellar position. This relation between velocity and radius can be explained by the free fall velocity with a central stellar mass of $\sim 0.3\ \Ms$. Hence, another interpretation of the blueshifted component is anisotropic mass infall from the envelope around SMM4B.

\section{ANALYSIS} \label{sec:analysis}
\subsection{Configuration of the SMM4B Outflow} \label{sec:conf}
\begin{figure*}[ht!]
\gridline{
\fig{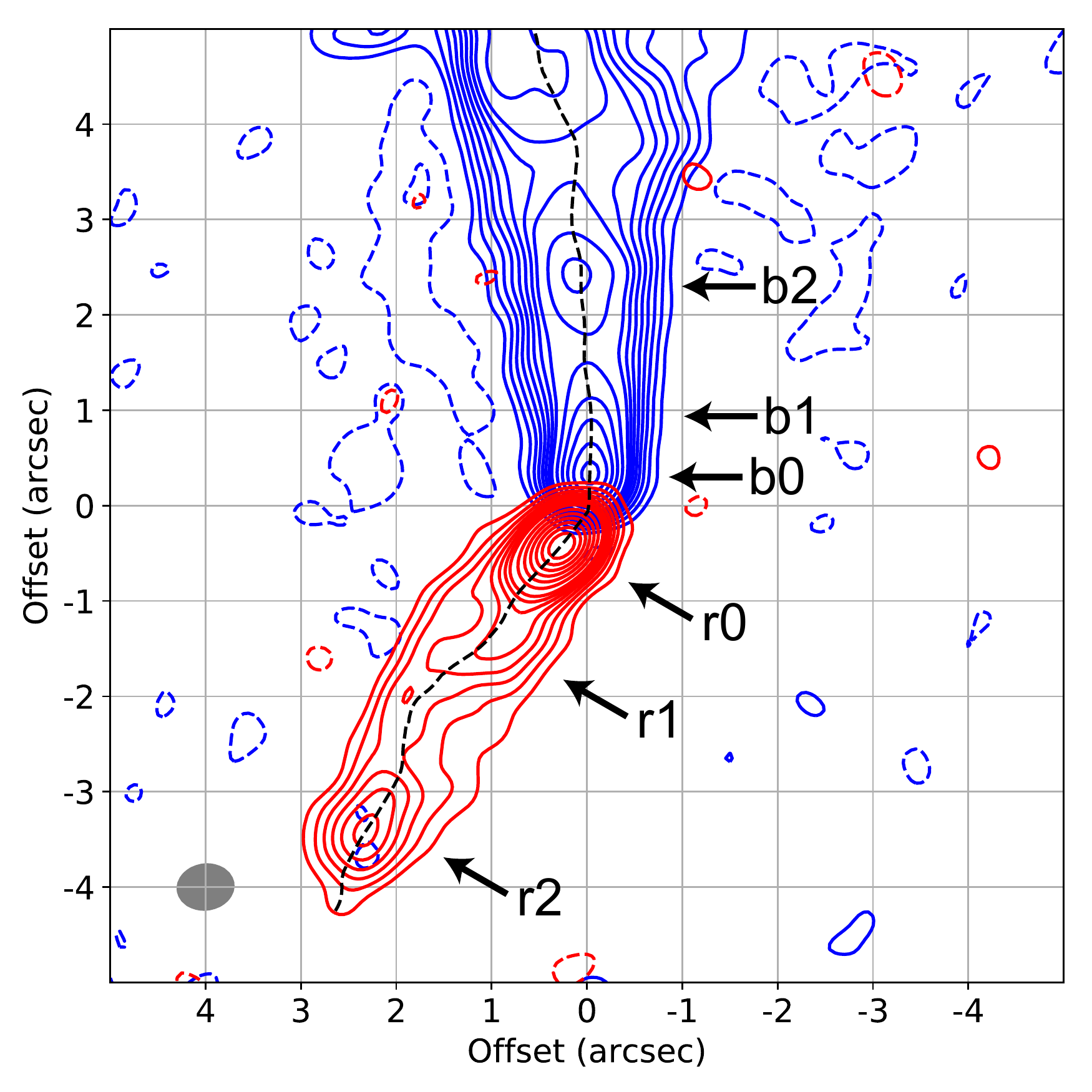}{0.3\textwidth}{(a)}
\fig{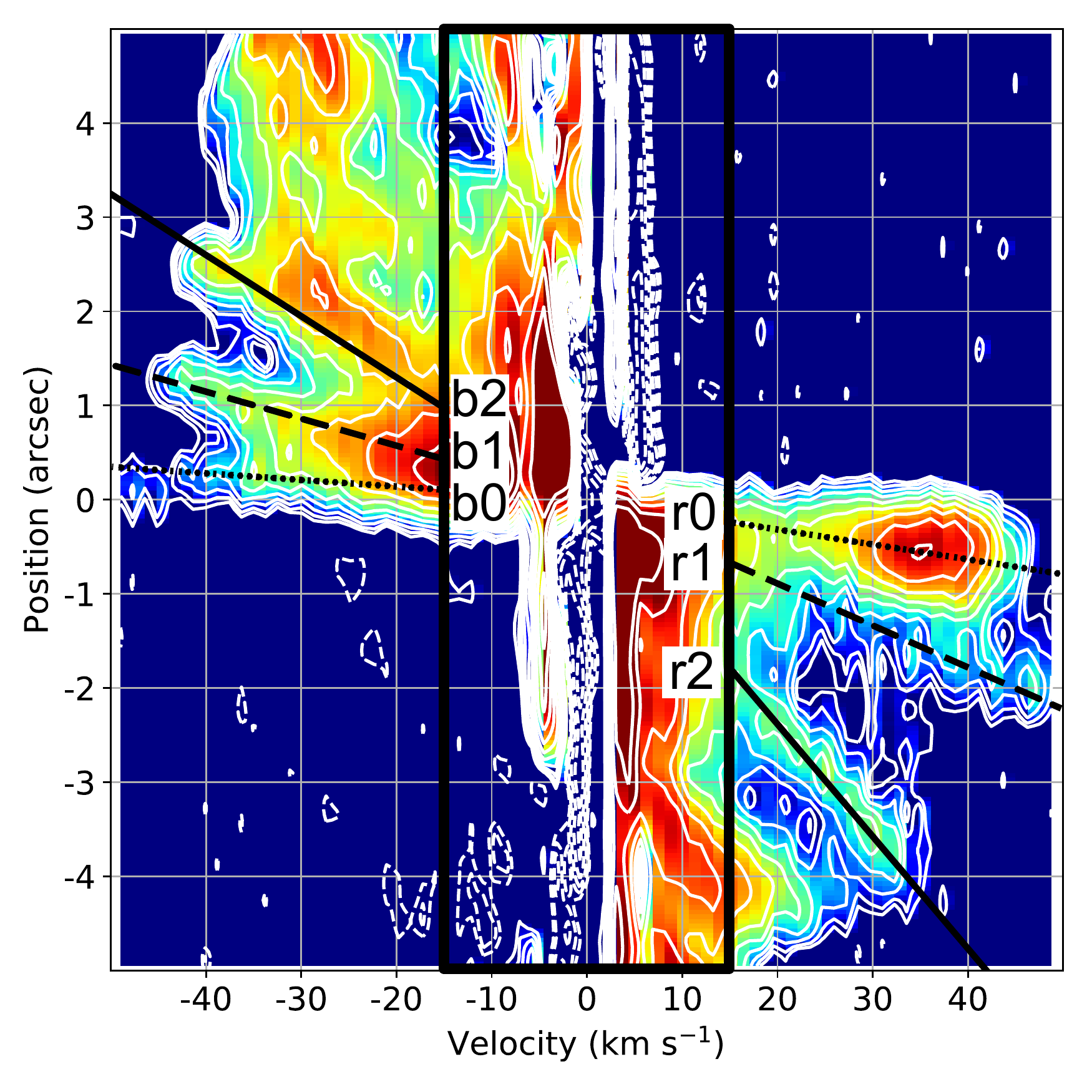}{0.3\textwidth}{(b)}
\fig{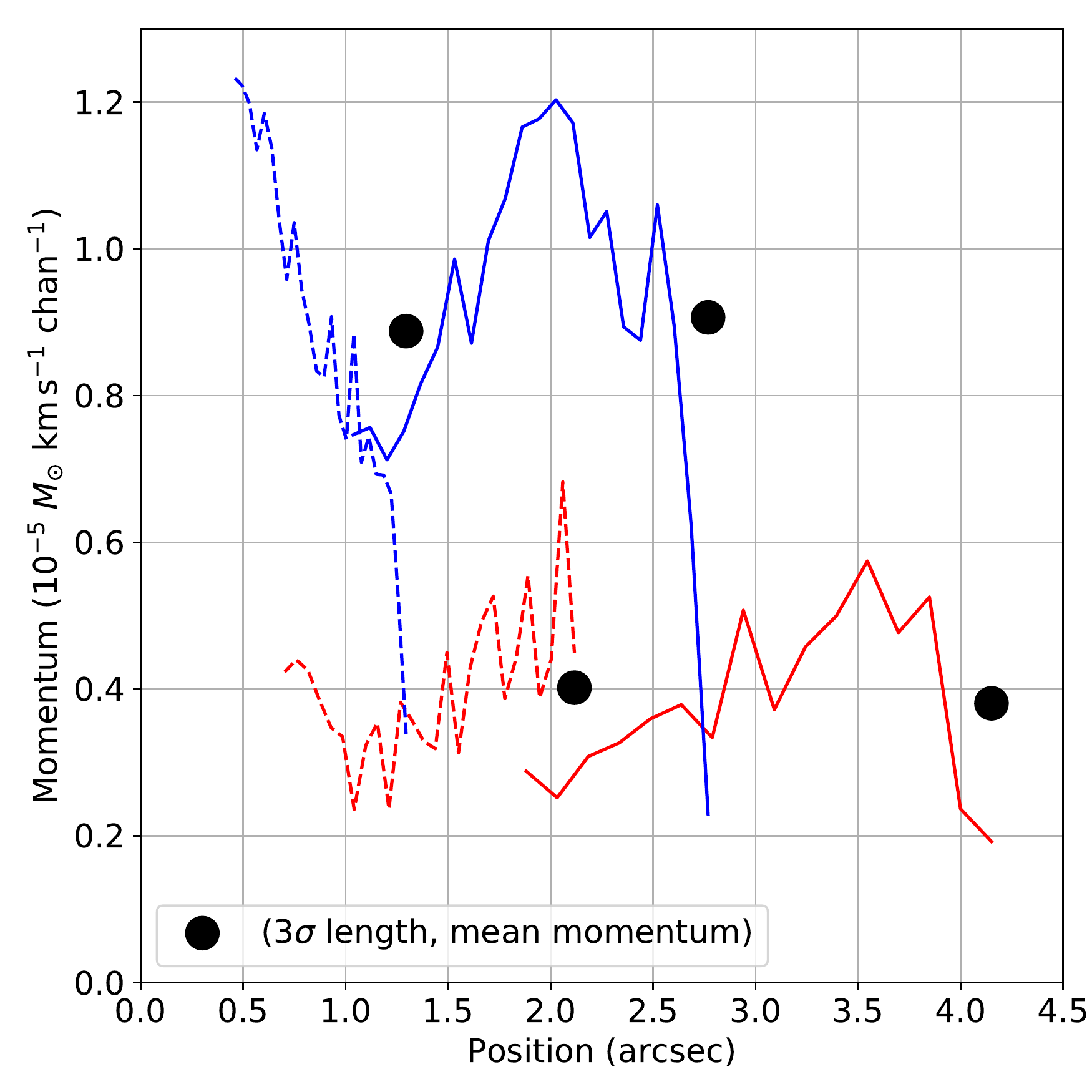}{0.3\textwidth}{(c)}
}
\caption{
(a) Integrated intensity maps of blue- and redshifted components of the $^{12}$CO emission. The integrated velocity range is $15\ \kms <|V-V_{\rm sys}|<50\ \kms$. Contour levels are from $3\sigma$ to $24\sigma$ in steps of $3\sigma$ and then in steps of $6\sigma$, where $1\sigma$ corresponds to $38\ \mJB~\kms$. The black dashed curve traces peaks of the integrated intensity maps. (b) Position-velocity diagram along the curve shown in panel (a). Contour levels are $2\sqrt{2},4,4\sqrt{2},8\dots \times \sigma$, where $1\sigma$ corresponds to $2.6\ \mJB$. Color also shows the same PV diagram in logarithmic scale. The rectangular frame denotes the velocity range $|V-V_{\rm sys}|<15\ \kms$. Solid and dashed lines trace Hubble-law motions derived by visual inspection; solid (dashed) lines are considered to be a pair (see the text in more detail). (c) Momentum profiles along the dashed and solid lines shown in panel (b). The width of cut is $0.55\arcsec$ (1 beam) along the positional direction. Blue and red lines show the profiles of the blue- and redshifted components, respectively. Circles signs denote mean momenta and $3\sigma$ lengths (positions) (see the text in more detail).
\label{fig:curved}}
\end{figure*}

To investigate the velocity structures of the $^{12}$CO outflow associated with SMM4B, we inspect the distribution of knots in the observed 3D space, i.e., right ascension, declination, and velocity.
First, an emission ridge in the 2D spatial domain is measured using $^{12}$CO maps integrated along the velocity direction. Figure \ref{fig:curved}a shows the blue- and redshifted integrated intensity maps of high velocity components; the integrated velocity range is $15\ \kms<|V-V_{\rm sys}|<50\ \kms$, where $V_{\rm sys}=7.9\ \kms$. The low velocity range is excluded because the $^{12}$CO emission shows complicated structures due to missing flux and/or self-absorption as seen in its channel maps (Figure \ref{fig:coch}), while the integrated velocity range includes all knotty structures that can be identified in the channel maps. The black dashed curve in Figure \ref{fig:curved}a traces the emission ridge of these integrated intensity maps, determined by Gaussian fitting along cuts across P.A.=$0\arcdeg$ and $150\arcdeg$ in the blue- and redshifted maps, respectively; the two directions are shown in Figure \ref{fig:12coso}a and \ref{fig:12coso}b with yellow dashed lines. The fitting includes only pixels with emission above $3\sigma$. Figure \ref{fig:curved}a also shows three pairs of knotty structures as pointed with arrows. We labeled them in the order of the distance from the central protostar as (b0, r0), (b1, r1), and (b2, r2). Although b0 and b1 appear not to be spatially separated in Figure \ref{fig:curved}a, these are identified in the channel maps (Figure \ref{fig:coch}) more distinctly. These pairs can also be identified in the PV diagram discussed below.

Second, the velocity structure of the $^{12}$CO emission is inspected in another 2D domain; the velocity axis versus the spatial trajectory derived above. Figure \ref{fig:curved}b shows that PV diagram in the 2D domain; the cut used to draw the PV diagram has a width of $0.55\arcsec$ (1 beam) in the direction perpendicular to the trajectory at each point (the black dashed curve in Figure \ref{fig:curved}a). The PV diagram exhibits Hubble-like flows as marked with solid, dashed, and dotted lines in Figure \ref{fig:curved}b. Those correspond to the three pairs of knotty structures identified in Figure \ref{fig:curved}a. These multiple Hubble-law lines indicate that mass ejection by the outflow is not continuous but episodic; the knot pair of b2 and r2 delineated by the solid lines in Figure \ref{fig:curved}b are older ejection than the other lines, since this pair reached the largest distance from the drinving source than the other knot pairs even though their terminal velocities are similar. Such episodic incidents of Hubble-like flows were also observed in previous studies of other protostellar molecular outflows \citep[e.g.,][]{pl2015}. We note that the Hubble-law line of r0 passes a strong secondary peak at $V-V_{\rm sys}\sim 35\ \kms$. Such a secondary peak is often seen in YSO jets \citep[e.g.,][]{ba1990,hi2010}. In contrast, line profiles of the other Hubble-like flows are more like wings than such a secondary peak; wings are often seen in outflows rather than jets. Jets and outflows can have different driving mechanisms; for example, jets are magnetically launched from the close vicinity of a central star, while outflows result from interaction between jets and surrounding material \citep{sh2000}. Their physical quantities, such as momentum, can thus be different. In addition to these physical differences, the pair (b0, r0) is geometrically much shorter than the other pairs, causing larger uncertainty in estimating inclination angles. For these reasons, we exclude this pair in the following analysis.

Third, the momentum is calculated along the Hubble-law lines (b1, r1) and (b2, r2) in Figure \ref{fig:curved}b. Figure \ref{fig:curved}c shows the momenta derived at each velocity channel as the products of fluxes, velocities, and a conversion factor between flux and mass \citep[e.g.,][]{ta2011,va2018} assuming optical thinness. This calculation assumes the local thermodynamic equilibrium (LTE) condition, the constant excitation temperature 30 K, and the $^{12}$CO abundance $X(^{12}{\rm CO})=2.7\times 10^{-4}$ \citep{la1994}. The adopted excitation temperature is typical in protostellar outflows in $^{12}$CO emission \citep[e.g.,][]{hi2010}, which ensures that the observed $^{12}$CO emission is optically thin. This calculation includes pixels within a width of $0.55\arcsec$ (1 beam) in the positional direction of Figure \ref{fig:curved}b. Since the cut used to derive Figure \ref{fig:curved}b also has the same width, the momentum calculation includes 1-beam-sized square areas in the 2D spatial domain. This momentum plot shows spatial changes of the flow momentum along each Hubble-law line. We also calculate the mean momentum along each flow line using the arithmetic mean without intensity weighting. We define the length of each Hubble-like flow as $3\sigma$ lengths, i.e., the distance from the central stellar position to the (longest) position where each Hubble-law line intersects a $3\sigma$ contour in the PV diagram Figure \ref{fig:curved}b. The pairs of the mean momenta and the $3\sigma$ lengths are plotted in Figure \ref{fig:curved}c with black circles.

Figure \ref{fig:curved}c shows three results: (1) the two blueshifted Hubble-like flows have similar momenta, (2) the two redshifted ones have similar momenta, and (3) the two pairs, blueshifted pairs and redshifted pairs, have significantly different momenta. One possible interpretation of these results is that all the four Hubble-like flows have the same intrinsic momentum, whereas different inclination angles between the blue- and redshifted pairs result in the difference of the ``projected" momenta as shown in Figure \ref{fig:curved}c. This interpretation is also consistent with the fact that the longer redshifted line (r2, solid red line in Figure \ref{fig:curved}c) is longer than the longer blueshifted line (b2, solid blue) and similarly the shorter redshifted line (r1, dashed red) is longer than the shorter blueshifted line (b1, dashed blue). From a viewpoint of momentum-conserving outflow models, such as the wind-driven-shell model, each part of a shell receives momentum from a wind/jet, which is consistent with the hypothesis that a pair of blue- and redshifted lobes receives a common momentum at a common length. Although it is also possible that the intrinsic momentum is different between the blue- and redshifted lobes of this outflow, we adopt the simplest hypothesis described above in the calculation below for the SMM4B outflow. This should be reasonable approximation because its bending suggests that the inclination angle difference can strongly affect the projected momentum in this system.

The inclination angles can be estimated from Figure \ref{fig:curved}c on the assumption discussed above. With a given inclination angle $i_b$ ($i_r$) of a blueshifted (redshifted) lobe ($0\arcdeg\leq i\leq 90\arcdeg$ and $0\arcdeg$ means the pole-on configuration), an intrinsic momentum, $p$, is observed as a projected momentum, $p_b=p\cos i_b$ ($p_r=p\cos i_r$), while an intrinsic length, $l$, is similarly observed as a projected length, $l_b=l\sin i_b$ ($l_r=l\sin i_r$). These are summarized as $l_b/l_r=\sin i_b/\sin i_r$ and $p_b/p_r=\cos i_b/\cos i_r$. These ratios are independent of the assumed excitation temperature and the conversion factor if those are the same in the blue and red lobes. 
With the derived mean momenta and the $3\sigma$ lengths, the inclination angles are calculated to be $i_b=36\arcdeg \pm 3\arcdeg$ and $i_r=70\arcdeg \pm 2\arcdeg$ for the two pairs (b1, r1) and (b2, r2). With the inclination angels and the apparent bending angle $\alpha ' =30\arcdeg$, the intrinsic bending angle in the 3D space, $\alpha$, can be written  as $\cos \alpha =\cos i_b \cos i_r +\sin i_b \sin i_r \cos \alpha '$ thus $\alpha\sim 40\arcdeg$. These inclination angles also provide an intrinsic momentum of $\sim 2$-$4\times 10^{-4}\ \Ms~\kms$ with the excitation temperature of 20-50 K for both older and newer pairs of mass ejection. Intrinsic lengths can also be estimated from the inclination angles to be $\sim 1000$ au and $\sim 2000$ au for the shorter and longer pairs, respectively.

\subsection{Continuum Visibility} \label{sec:vis}
\begin{figure}[ht!]
\epsscale{1}
\plotone{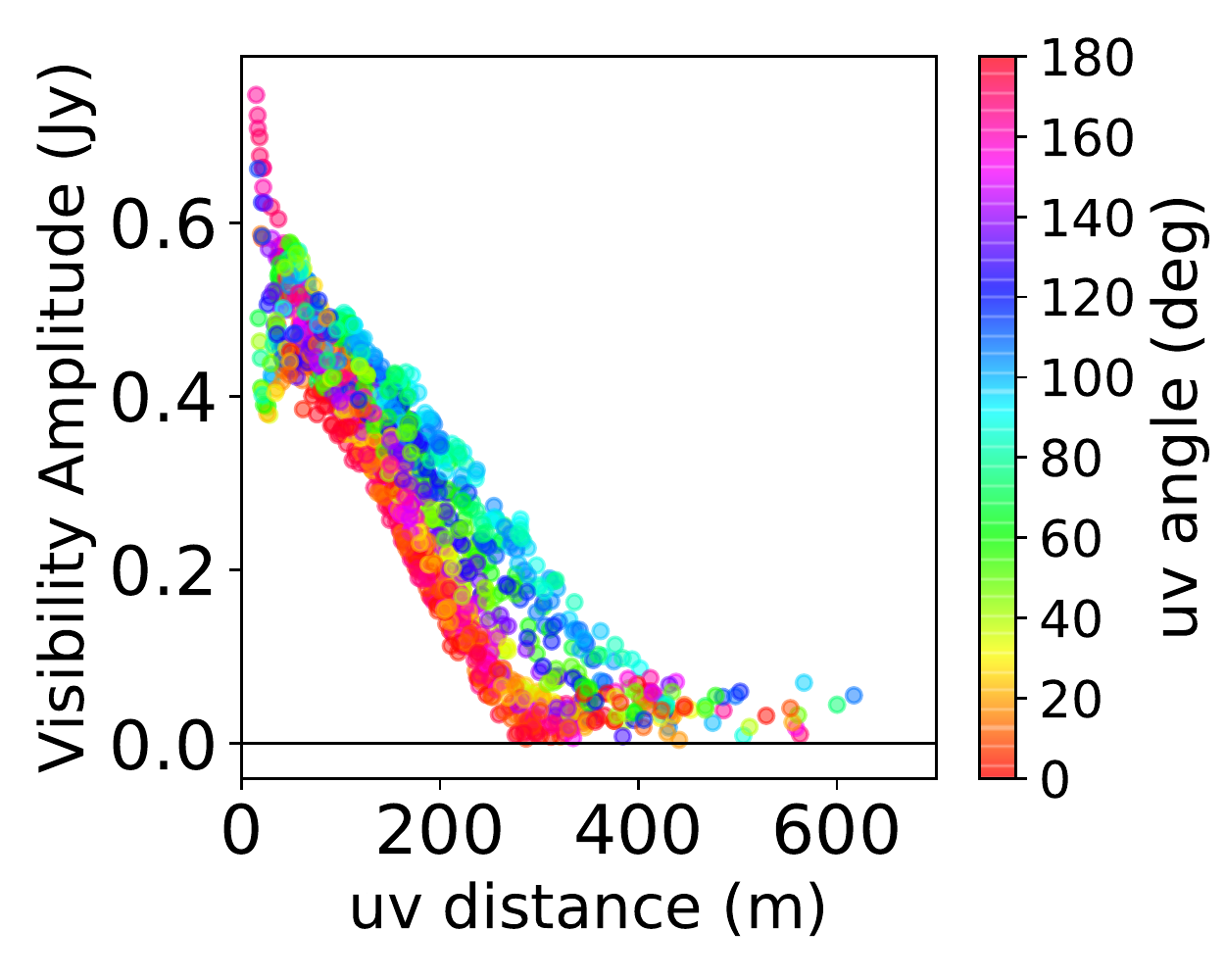}
\caption{Plots of the observed continuum visibilities in 1.3 mm, including all components associated with both SMM4A and SMM4B. The $uv$ distance of 1 m corresponds to 0.78 k$\lambda$. The visibility data were averaged over scans and thus each point corresponds to each pair of antennas. Error bars of the visibility amplitude are $\lesssim 2.5\ {\rm mJy}$. The phase reference position is the SMM4A position. 
Color denotes relative position angles in the $uv$ plane from the major-axis P.A.=145$\arcdeg$.
\label{fig:visobs}}
\end{figure}

Young stellar objects show remarkable structures depending on their evolutionary phases, growing disks out of spherical or flattened envelopes. To reveal such structures related to evolutionary phases in SMM4A and SMM4B, the observed continuum visibilities are inspected in this subsection in detail, which unveil intriguing aspects of the continuum structures more clearly than from the image inspection (Section \ref{sec:cont}). The visibility plots shows in Figure \ref{fig:visobs} are derived from the averaging over all scans for each pair of antennas.

Since the total flux density of the continuum emission in SMM4A is $\gtrsim 3$ times larger than that in SMM4B, the visibility distribution is expected to be dominated by flux from SMM4A. In fact, Figure \ref{fig:visobs} shows that the 2D visibility amplitude distribution is extended in the perpendicular direction to the major axis of SMM4A derived in the image domain.
In other words, the profiles in Figure \ref{fig:visobs} are narrower at angles closer to the major axis ($0\arcdeg$ or $180\arcdeg$ in Figure \ref{fig:visobs}). Furthermore this figure also shows that radial profiles of the visibility amplitude at position angles close to the major axis have a null point at a $uv$-distance of $\sim 300$ m. Radial profiles at angles close to the minor axis also exhibit a possible null point at a greater $uv$-distance. 

\begin{figure}[ht!]
\epsscale{1.2}
\plotone{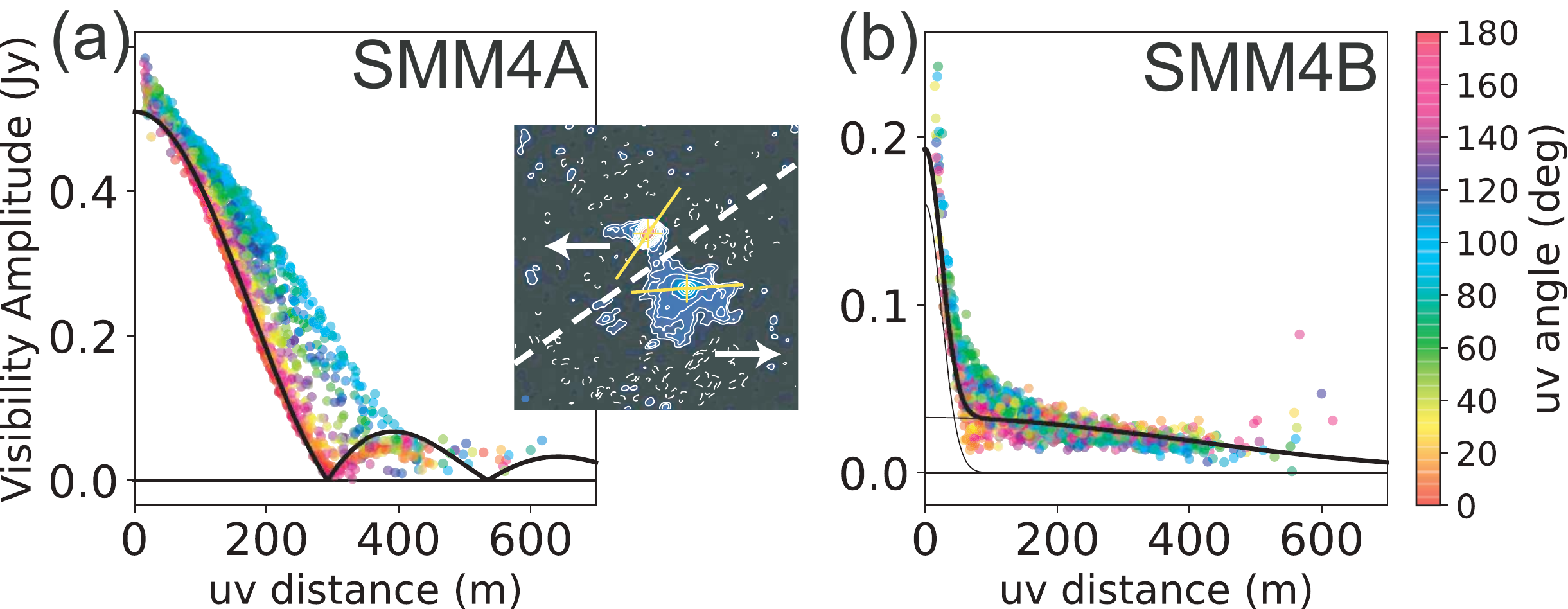}
\caption{Same as Figure \ref{fig:visobs} but for visibilities of SMM4A (a) and SMM4B (b). Color shows the relative position angles from the major axes of (a) SMM4A, P.A.$=145\arcdeg$ and (b) SMM4B, P.A.$=94\arcdeg$. The CLEAN components were divided into ones associated with SMM4A and ones associated with SMM4B by the dashed line shown in the inset, which is the same as Figure \ref{fig:cont}. The curve in panel (a) denotes Fourier transform of a uniform (boxcar) disk with a radius of 240 au. The thick curve in panel (b) denotes Fourier transform of a double Gaussian profile with radii (HWHMs) of 590 au and 56 au, while the thin curves denote each Gaussian component. 
\label{fig:cc}}
\end{figure}

The null point is expected to be associated with SMM4A because of the flux difference between SMM4A and SMM4B. To make it clear and to investigate structures of SMM4B as well, the visibility data are analyzed in more detail. A simple way to separate the visibility data between SMM4A and SMM4B is removing the CLEAN components of one source from the observed visibility data. The CLEAN components in the image domain were transformed to the visibilities by synthetic observations on the same conditions as those of our observations. Synthetic observations were performed using the CASA task $simobserve$. Then, the visibilities of the CLEAN components were subtracted from the observed visibilities. Figure \ref{fig:cc}a shows the observed visibilities minus the visibilities derived from the SMM4B CLEAN components, while Figure \ref{fig:cc}b shows ones derived by subtracting the SMM4A CLEAN components. The dashed line in the inset shows a boundary used to split the CLEAN components, where the upper-left and lower-right parts are regarded as components associated with SMM4A and SMM4B, respectively.

Figure \ref{fig:cc}a and \ref{fig:cc}b show amplitude profiles of the split visibilities. The profiles at $uv$-angles $\sim 0\arcdeg$ and $\sim 180\arcdeg$ are narrower than those along the other directions in Figure \ref{fig:cc}a and \ref{fig:cc}b. This is consistent with the result from the Gaussian fitting in the image domain. As expected from the original visibility data, the visibility amplitude profiles of SMM4A have a null point at a $uv$-distance of $\sim 300$ m at position angles close to the major axis, suggesting the presence of a disk-like structure with a sharp edge. To confirm it and measure a typical size of the structure, we fit the visibility amplitude profile with the Fourier transform of a uniform (boxcar) disk in order to focus on the remarkable geometry, sharp edge. The Fourier transformation of such a uniform disk is described as $2A_0 J_1(1.22 \pi \beta /\beta_1)/(1.22 \pi \beta /\beta_1)$, where $A_0$, $J_1$, $\beta$, and $\beta_1$ are the visibility amplitude, the Bessel function of order 1, the $uv$-distance, and the first null point, respectively. The curve in Figure \ref{fig:cc}a shows the best-fit result with this function, $(A_0,\beta _1)=(500\pm 5\ {\rm mJy},293\pm 3\ {\rm m})$, derived by fitting visibilities in the directions around the major axis ($\pm 10\arcdeg$); this first null point corresponds to a radius of 240 au in the image domain. The deviation from the best-fit curve at $\sim$400 m implies that the edge of the true intensity profile is not so sharp as that of the uniform disk. In addition, the amplitude at very short $uv$-distances ($\lesssim 50$ m) is slightly higher than the best-fit curve, which may correspond to extensions seen in the image domain (Figure \ref{fig:cont}).

Not only the SMM4A profiles but also the SMM4B profiles (Figure \ref{fig:cc}b) show a remarkable feature, a combination of compact and extended components.
The two components are fitted by a double Gaussian fitting using visibilities in the directions around the major axis ($\pm10\arcdeg$) as done in the fitting to the SMM4A visibilities. The curves in Figure \ref{fig:cc}b show the best-fit results: the extended and compact components have the peak amplitudes of $150\pm 9$ mJy and $33\pm 1$ mJy, respectively. The sizes of these components in HWHMs (radii) are $29\pm 1$ m and $450\pm 50$ m, which correspond to radii of 590 au and 56 au, respectively, in the image domain. The extended component can also be seen in the image domain (Figure \ref{fig:cont}).

\subsection{C$^{18}$O abundance} \label{sec:abundance}
On the basis of the continuum structures described above, we can calculate the column density of H$_{2}$ molecules along each line of sight with the following equation:
\begin{eqnarray}
I_{\rm cont.}=(B_{\nu}(T)-B_{\nu}(T_{\rm bg}))(1-e^{-\tau _{\rm dust}}),
\end{eqnarray}
where $I_{\rm cont.}$, $B_{\nu}$, $T$, $T_{\rm bg}=2.73$ K, and $\tau _{\rm dust}$ are the observed continuum intensity, the Planck function, the dust/gas temperature, the background temperature, and the dust optical depth, respectively.
The dust opacity law, gas and dust temperature, and the gas-to-dust mass ratio are assumed to be $\kappa (850\ \micron)=0.035\ {\rm cm}^{2}~{\rm g}^{-1}$ \citep{an.wi2005}, $\beta =1$, $T=20$ K \citep{leKI2014}, and 100, respectively. In addition, the mean molecular weight 2.37 is adopted to change the mass column density of gas to the number column density of $H_{2}$ molecules.

Similarly, the observed line intensity before continuum subtraction is
\begin{eqnarray}
I_{\rm line}=(B_{\nu}(T)-B_{\nu}(T_{\rm bg}))(1-e^{-(\tau _{\rm gas}+\tau _{\rm dust})}),
\label{eq:gas}
\end{eqnarray}
where $I_{\rm line}$ and $\tau _{\rm gas}$ are the observed line intensity and the optical depth of the molecular line. 
By applying Equation (\ref{eq:gas}) to the observed C$^{18}$O emission above the $3\sigma$ level at each channel, we can derive the optical depth at each channel. We assume LTE and the same gas temperature as the dust. The C$^{18}$O column density at each channel is derived from the optical depth at each channel, and the total column density is derived by summing up the column density at each channel. The fractional abundance of C$^{18}$O relative to H$_2$, $X$(C$^{18}$O) is then derived from the ratio of the two column densities.

\begin{figure}[ht!]
\epsscale{1}
\plotone{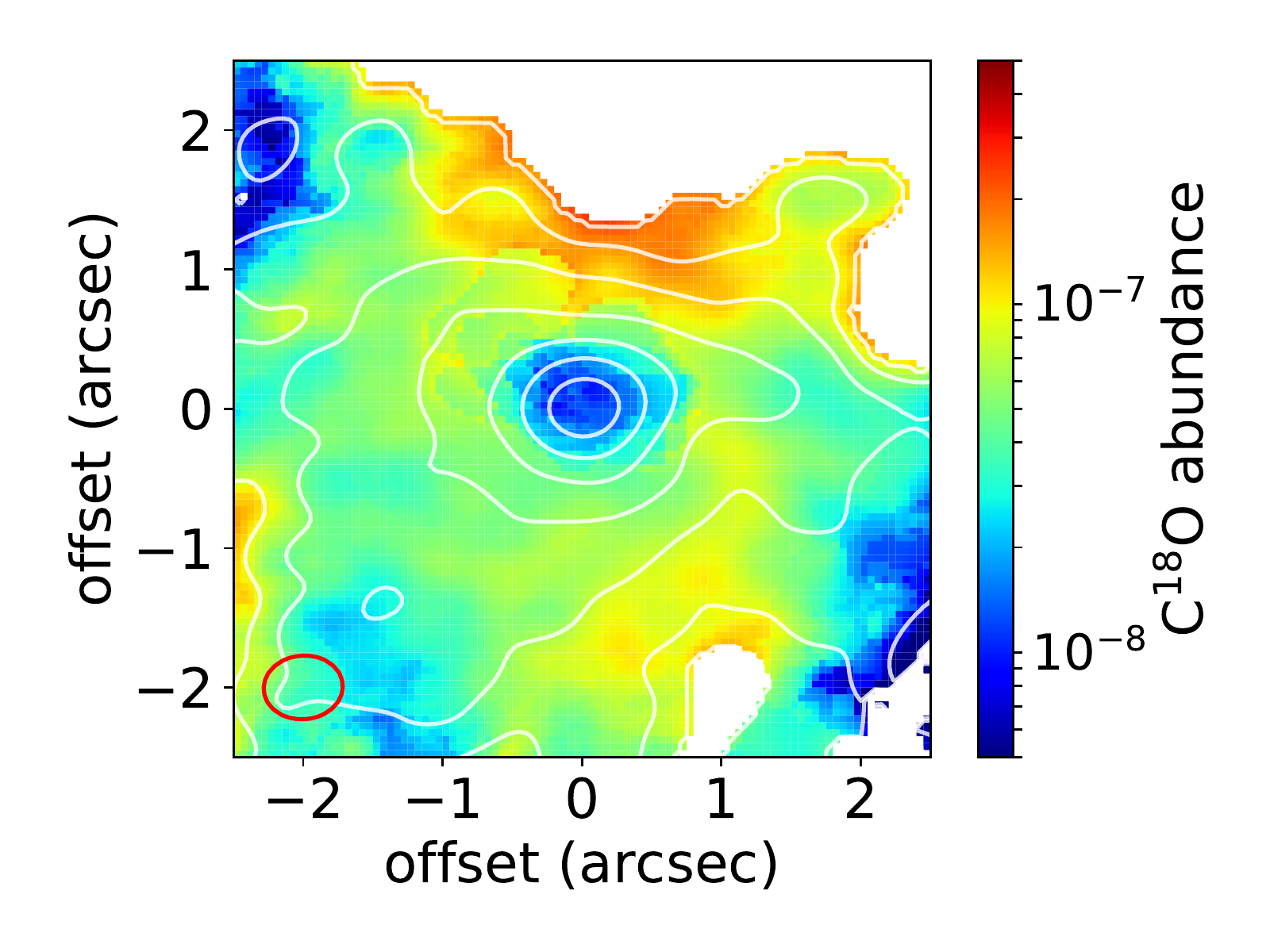}
\caption{Map of C$^{18}$O abundance estimated from the observed C$^{18}$O and continuum emission. Coordinates are offsets from the SMM4B position, where $1\arcsec$ corresponds to 429 au. Contours and synthesized beam (red ellipse) are the same as those in Figure \ref{fig:cont}. Pixels are masked if the continuum intensity is below the $3\sigma$ level.
\label{fig:dep}}
\end{figure}

Figure \ref{fig:dep} shows the spatial distribution of the derived C$^{18}$O abundance. The abundance is $\lesssim 10^{-8}$ at the SMM4B position, which is $\gtrsim 50$ times lower than one in the interstellar medium (ISM) \citep[e.g., $5\times 10^{-7}$,][]{la1994,wi.ro1994}.
On the other hand, Figure \ref{fig:dep} shows that the abundance in the surrounding region is $\sim 5\times 10^{-8}$ on a $\sim 1500$ au scale, which is consistent with measurement on a $\sim 5000$ au scale \citep{d-c2010}. The derived optical depth of the C$^{18}$O line is $\lesssim 1$. The abundance would be slightly lower if we assumed purely optically thin emission. In addition, higher temperatures result in higher C$^{18}$O abundances because of smaller population in the upper level of the transition $J=2-1$; 15-40 K provides $X({\rm C}^{18}{\rm O})=0.3$-$3\times 10^{-8}$ at the SMM4B position. The C$^{18}$O abundance is thus at least 10 times smaller than the ISM value in this temperature range.

We are unable to carry out the same analysis for SMM4A as the C$^{18}$O line shows negative intensities probably due to foreground absorption.


\section{DISCUSSION} \label{sec:discussion}
\subsection{Possible Mechanisms of the Bending Outflow} \label{sec:bend}
In Section \ref{sec:conf}, we found that the SMM4B $^{12}$CO outflow bends at an angle $\sim 40\arcdeg$. Since similar bending outflows have been reported in previous observations of protostars \citep[e.g.,][]{ch2016,to2016,ye2017,aso2017b}, it is worth considering possible mechanisms of this apparently common outflow bending. In the case of SMM4B, one of the simplest possibilities is the orbital motion between SMM4A and SMM4B. It is, however, not likely, even if the two protostars are gravitationally bound, because their projected separation corresponds to $\sim 5\arcsec$ (2100 au). Such a wide orbit only yields an orbital velocity of $\lesssim 1\ \kms$ with a central stellar mass of $\sim 1\ \Ms$, for example, which is negligible compared to the outflow velocity of several tens $\kms$. The possible orbital motion, therefore, cannot explain the bending of the SMM4B outflow. It is also unlikely either that SMM4B itself is a unresolved close binary system having a separation smaller than several tens au. In the case of such a close binary system, the orbital motion produces a wiggling pattern in the outflow lobes \citep{ma.ra2002,wu2009}, rather than a large bending angle. Dynamical interaction with dense gas also has a potential for bending the outflow \citep{um1991}. The Serpens Main region was observed with CARMA at an angular resolution of $\sim 7\farcs 5$ in a dense gas tracer, HCN $(1-0)$ \citep[Figure 3 in][]{leKI2014}. HCN $(1-0)$ has a critical density $\sim 2$ orders of magnitude higher than that of C$^{18}$O $J=2-1$, although it could trace less dense regions depending on optical depth. Their result shows strong compact emission in the vicinity of SMM4A and relatively strong emission extended from the north to the northwest of SMM4A and SMM4B. This configuration of dense gas is not likely either to bend the blue lobe of the SMM4B outflow from the northwest to the north as discussed in \citet{um1991}. Similarly it is not likely that the SMM4A outflow changes the direction of the SMM4B blue lobe as observed, since SMM4A is located in the northeast of SMM4B and its outflow goes to the north. On the other hand, outflows on 0.1 pc scales are identified in the whole of Serpens Main \citep{da1999,gr2010} by JCMT observations; those outflows could push the SMM4B outflow from the western side, while such large-scale outflows would be resolved out in our ALMA observations.

Lastly, we discuss an idea of electromagnetic interaction, which is introduced to explain a similar outflow bending in the proto-binary system NGC 1333 IRAS 4A \citep{ch2016}. They discussed the action of the Lorentz force between magnetic fields associated with one outflow and the current of the other outflow on the basis of submillimeter polarization observations. In the case of SMM4B, the position angle of the blue lobe, i.e., current can rotate counterclockwise if the magnetic field acting on the blue lobe is strong enough along the line-of-sight from the near side to the far side. The calculation in \citet{ch2016} is based on the ideas by \citet{fe.zi1998}: the Lorentz force can be written as $jB=\rho V^2 /R_{\rm \kappa}$, where $j$, $B$, $\rho$, $V$, and $R_{\kappa}$ are the current density carried by an outflow, the magnetic field flux, the mass density, the outflow velocity, and the curvature radius, respectively. The curvature radius $R_{\kappa}$ is written as $R_{\kappa}\sim L/(2\sin \alpha)$, where $L$ and $\alpha$ are the outflow length and the bending angle, respectively; this relation is different from the one adopted in \citet{fe.zi1998} because we define $L$ as the linear distance from the central protostar to an edge of the outflow and $\alpha$ is not necessarily small here. Since $j=I/(\pi R^2)$ and $\rho =M/(\pi R^2L)$, where $I$, $R$, and $M$ are the current, the outflow radius, and the outflow mass, respectively, $B=2\sin \alpha MV^2 /(IL^2)$. The current $I$ can be scaled by typical values as $I=10^{11}\ {\rm A}\ (R/70\ {\rm au})^2(n/100\ {\rm cm}^{-3})(V/300\ \kms)$ \citep{fe1995} as done in \citet{ch2016}, where $n$ is the number density. With the typical molecular weight $m=2.37m_{\rm H}$, i.e. $n=\rho /m$, the required magnetic flux density can be calculated as $B=0.55\ {\rm mG}\ \sin \alpha (\tan i_b)^{-1} (V'/10\ \kms)(L'/1000\ {\rm au})^{-1}$, where $i_b$, $V'$, and $L'$ are the inclination angle of the blue lobe, its line-of-sight velocity, and its projected length, respectively. The other factors cancel, resulting in this simple dependency, because of the current scaling which fixes a certain efficiency of the Lorentz force, such as the ionization rate. As a result, stronger magnetic fields are required to bend an outflow with higher velocities and shorter lengths. If we adopt $(\alpha,i_b,V',L')\sim (40\arcdeg,36\arcdeg,40\ \kms,1000\ {\rm au})$ for the blue lobe of the SMM4B outflow, the required magnetic flux density is $\sim 2\ {\rm mG}$. Although these measured quantities have uncertainties, the calculated magnetic flux density is likely on the order of mG, which is reasonable around protostars \citep{cr2010,gi2006,fa2008,st2013}. 
It is quantitatively plausible, therefore, that Lorentz force acting between the two outflows may bend the SMM4B outflow by the measured bending angle. 

In summary, two possibilities for bending the SMM4B outflow remain through the discussion above: one is dynamical interaction with large scale outflows and the other is electromagnetic interaction with mG-order magnetic fields. In addition, the fact that the SMM4B outflow appears to bend in the close vicinity of the central protostar might imply a relation between the bending and a disk-scale structure, such as a warped disk, as suggested by theoretical simulations \citep{ma2017}. Such scales, however, cannot be spatially resolved with the present observations. It is thus crucial to observe polarization or large/small-scale structures in order to directly verify these possibilities in the future.

\subsection{Evolutionary Phases} \label{sec:evo}
Both of the two protostars SMM4A and SMM4B are faint at 70 $\micron$ and shorter wavelengths, and their $T_{\rm bol}$ and $L_{\rm bol}/L_{\rm submm}$ indicate that they are classified as Class 0. On the other hand, our ALMA observations have revealed different physical properties between the two protostars in $^{12}$CO outflows and dust structures traced by 1.3 mm continuum. These differences may help us to distinguish evolutionary phases of the two protostars in more detail than the classification based on their SED, Class 0. We previously presented ALMA observations of another Class 0 protostar, SMM11, in the same star forming cluster Serpens Main \citep{aso2017b}. Its bolometric temperature $\sim26$ K, bolometric luminosity $\lesssim 0.91\ \Ls$, and internal luminosity $\lesssim 0.04\ \Ls$ are lower than those of SMM4A and SMM4B, if the two have half of the luminosities of SMM4. The C$^{18}$O freeze-out is also observed in SMM11 as is in SMM4B. The evolutionary phases of the three protostars are discussed in this subsection.

Our analysis of the continuum visibility profile of SMM4A suggests the presence of a disk-like structure with a radius of $\sim240$ au, while that of SMM4B suggests the presence of a compact disk with a radius of $\sim 56$ au embedded in an extended envelope. In contrast, the continuum visibility profile of SMM11 suggests a spherical envelope without an embedded disk nor a compact component \citep{aso2017b}, although the bipolar $^{12}$CO outflow indicates the existence of an very small, unresolved disk. These differences may imply the growth of protostellar disks from SMM11 to SMM4B and SMM4A in this order.

The disk growth scenario is also consistent with the difference in the SO emission between SMM4A and SMM4B: SMM4B is associated a strong compact component, while SMM4A has no such component. More energetic infall shock is anticipated in earlier evolutionary phases because of smaller disks. 
Mass infall onto the compact component (possible small disk) can cause shock waves that produce SO emission, while no shocked region in the vicinity of SMM4A suggests that the mass infall onto the SMM4A disk is less energetic than that onto the SMM4B disk. 

Their $^{12}$CO outflows and C$^{18}$O abundances also support this order of evolution as follows. SMM4A has a fan-shaped outflow (opening angle $\sim 90\arcdeg$) with low line-of-sight velocities, $\sim 10\ \kms$, while SMM4B has a very collimated outflow (opening angle $\lesssim 30\arcdeg$) with high line-of-sight velocities, a few $10\ \kms$; SMM11 has a relatively collimated outflow (opening angle $\sim 30\arcdeg$), whose inclination-corrected velocity is estimated to be a few $10\ \kms$ \citep{aso2017b}. Theoretical simulations predict that an outflow opening angle is more widened in later evolutionary phases \citep{ar.sa2006,ma.ho2013}, suggesting that SMM4A is older than SMM4B and SMM11. The C$^{18}$O abundance at the SMM4B position is lower than the ISM value by a factor of $\gtrsim 50$, while it is $\sim 1000$ times lower than the ISM value in SMM11. The low C$^{18}$O abundance indicates that the temperature of the gas envelopes is lower than the CO freeze-out temperature of 20-30 K in the central hundreds au regions of these sources. In contrast, the high brightness temperature, $\sim 18$ K, of continuum emission at SMM4A suggests a higher temperature in SMM4A, although C$^{18}$O abundance cannot be estimated around SMM4A because of the strong continuum emission. Theoretical models predict the increase of temperature along with evolution on such a spatial scale in the Class 0 phase \citep{ma.in2000,ai2012}. For these reasons, our results suggest that SMM4A is the most evolved among the three Class 0 protostars, followed by SMM4B and SMM11 in this order. 

Previous observations toward different regions also identified non-coeval Class 0 protostars formed in the same cores: VLA 1623A and 1623B \citep{mu.la2013} in $\rho$ Ophiuchus, B1-bN
and B1-bS \citep{hi.li2014} in Barnard 1, and IRAS 16293-2422 Source A and Source B in Ophiuchus \citep{ta2007}, for example. Since those protostellar pairs are both in the Class 0 phase, their age differences must be on the order of 0.1 Myr \citep{an.mo1994,ev2009,du2015} or shorter. Such a small age difference may not be evident in the later evolutionary stages, but the slight difference could be clearly identified in the earlier evolutionary stages when the evolutionary time scale is much shorter. In this context, the large disk in SMM4A may suggest rapid disk formation in an early phase of star formation, although other effects, such as initial conditions and environment, must also be taken into account.
The combination of continuum and molecular lines at millimeter wavelengths allow us to differentiate such phases in rapid evolution of deeply embedded protostars in detail.

\section{CONCLUSIONS} \label{sec:conclusions}
We have observed the submillimeter continuum condensation SMM4 in Serpens Main with ALMA during its Cycle 3 at angular resolutions of $\sim 0.55\arcsec$ (240 au) in the 1.3 mm continuum, the $^{12}$CO $J=2-1$, SO $J_N=6_5-5_4$, and C$^{18}$O $J=2-1$ lines. The main results are summarized below.

1. The high resolution continuum image reveals two compact sources, SMM4A and SMM4B, as well as an extended structures around SMM4B. SMM4A has a high brightness temperature of 18 K, while SMM4B has much lower brightness temperature of 2 K. The continuum visibilities of SMM4A suggest presence of a disk-like structure with a sharp edge at $r\sim 240$ au, while those of SMM4B suggest presence of a compact component, or a possible small disk, with a radius of 56 au.

2. The $^{12}$CO emission traces a fan-shaped blueshifted unipolar outflow associated with SMM4A. The outflow from SMM4B is bipolar and collimated. The axes of the blue- and redshifted lobes of the SMM4B outflow are misaligned by $30\arcdeg$. The SO emission traces shocked regions at the edges of the SMM4B outflow lobes and in the vicinity of SMM4B. The PV diagram of the $^{12}$CO along the lobes shows two pairs of linear (Hubble-law) features. By assuming that these two pairs of Hubble-law features have same intrinsic momenta, the inclination angles of the blue and red lobes were estimated to be $i_b\sim 36\arcdeg$ and $i_r\sim 70\arcdeg$, respectively, from the line of sight. The misalignment between blue and red lobes in three dimensional space was estimated to be $\sim 40\arcdeg$.

3. The origin of the bending of the SMM4B outflow could be 1) dynamical interaction with 0.1-pc-scale outflows, 2) electromagnetic interaction between current due to the outflow and magnetic fields with mG-order magnetic flux densities, and/or 3) smaller scale structures, such as a warped disk, which cannot be spatially resolved in our observations.

4. The C$^{18}$O line shows an absorption feature against the bright continuum emission at SMM4A, while it is observed as an emission feature at SMM4B. The C$^{18}$O around SMM4B mainly traces an infalling and rotating envelope. The C$^{18}$O abundance at the continuum peak of SMM4B was found to be $\sim  50$ times smaller than that of a typical ISM value, while the abundance on a $\sim 1500$ au scale is consistent with large scale observations of this region.

5. We have compared the evolutionary phases of SMM4A, SMM4B, and another Class 0 protostar, SMM11, in the same region. SMM4B having a collimated outflow, a compact continuum source, and mass accretion traced in the SO line is considered to be in an earlier evolutionary phase than SMM4A with a higher brightness temperature, a fan-shaped outflow, and a large disk-like structure. SMM11 shows a bolometric temperature, a bolometric luminosity, and an internal luminosity lower than those of SMM4A and SMM4B. In addition, the C$^{18}$O abundance of SMM11 is a few tens times smaller that of SMM4B. The continuum emission from SMM11 shows a spherical envelope without a detectable disk nor a compact component. For these reasons, SMM11 is considered to be even younger than SMM4B. It is likely that SMM4A is in the most evolved phase followed by SMM4B and SMM11 in this order.

\acknowledgments
This paper makes use of the following ALMA data: ADS/JAO.ALMA2015.1.01478.S (P.I. Y. Aso). ALMA is a partnership of ESO (representing its member states), NSF (USA) and NINS (Japan), together with NRC (Canada), NSC and ASIAA (Taiwan), and KASI (Republic of Korea), in cooperation with the Republic of Chile. The Joint ALMA Observatory is operated by ESO, AUI/NRAO and NAOJ.
We thank all the ALMA staff making our observations successful. We also thank the anonymous referee, who gave us invaluable comments to improve the paper.
Data analysis were in part carried out on common use data analysis computer system at the Astronomy Data Center, ADC, of the National Astronomical Observatory of Japan.
Y.A. acknowledges a grant from the Ministry of Science and Technology (MoST) of Taiwan (MOST 106-2119-M-001-013).
N.H. acknowledges a grant from the Ministry of Science and Technology (MoST) of Taiwan (MOST 106-2112-M-001-010).
S.T. acknowledges a grant from JSPS KAKENHI Grant Number JP16H07086 in support of this work. This work was supported by NAOJ ALMA Scientific Research Grant Numbers 2017-04A.

%

\vspace{5mm}
\facilities{ALMA}


\software{CASA, MIRIAD}


\appendix
\section{Channel Maps}
\label{sec:app_ch}
In this appendix, we present the channel maps of the observed lines: $^{12}$CO $J=2-1$, SO $J_N=6_5-5_4$, and C$^{18}$O $J=2-1$. The velocity resolution is adjusted in each line to show a whole velocity range where the line is detected. Systemic velocities of SMM4A and SMM4B are estimated to be $\sim 7.46$ and $7.86\ \kms$, respectively, in Section \ref{sec:c18o}.

\begin{figure}[ht!]
\epsscale{1.2}
\plotone{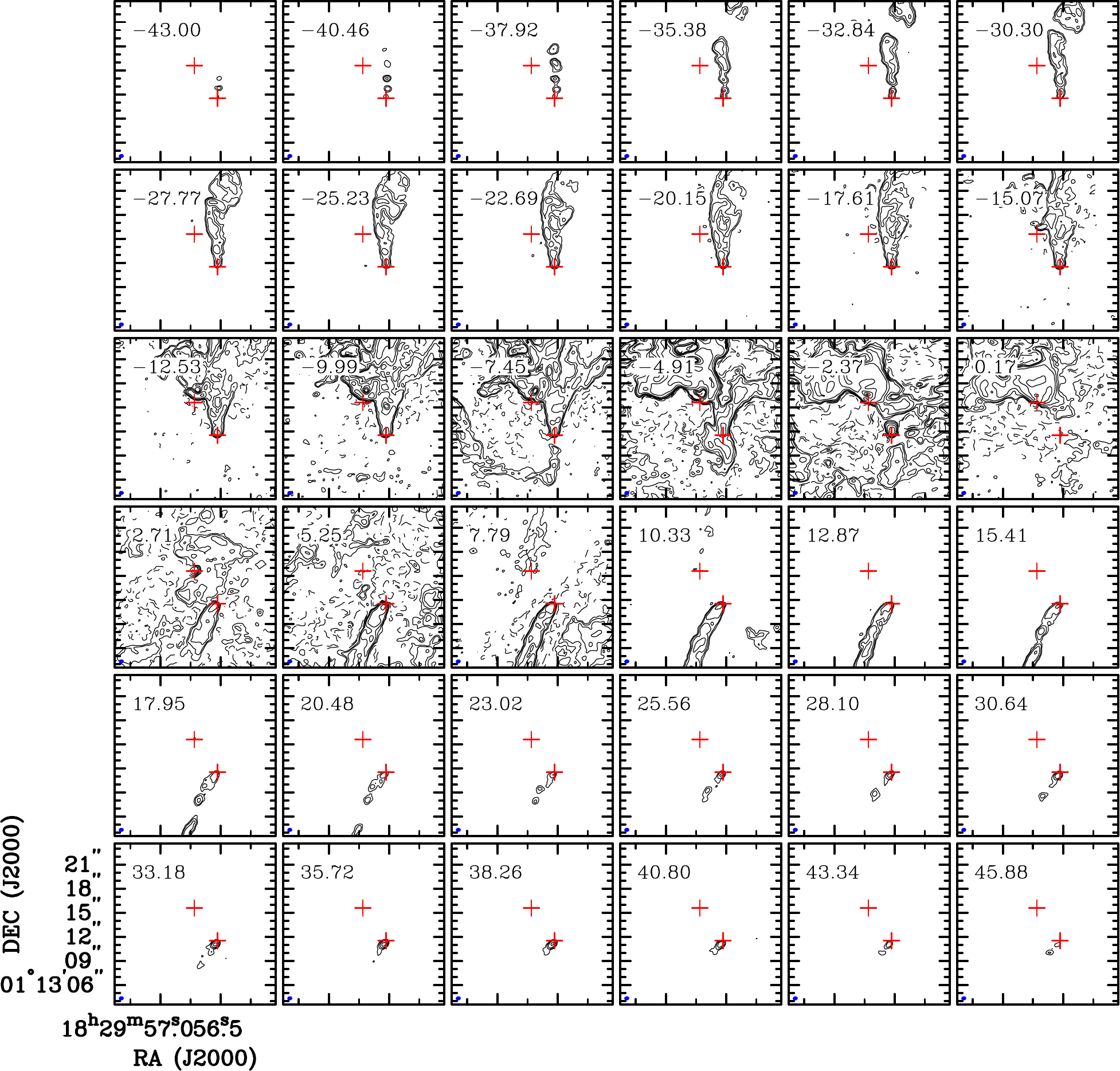}
\caption{Channel maps of the $^{12}$CO $J=2-1$ emission line. The velocity resolution is $2.54\ \kms$. Contour levels $5,10,20,40,\dots \sigma$, where $1\sigma$ corresponds to $2.6\ \mJB$. Two red plus signs and a blue ellipse in each panel are the continuum peak positions and the ALMA synthesized beam, $0\farcs 61\times 0\farcs 50,\ {\rm P.A.}=-82^{\circ}$, respectively.
\label{fig:coch}}
\end{figure}

\begin{figure}[ht!]
\epsscale{1.2}
\plotone{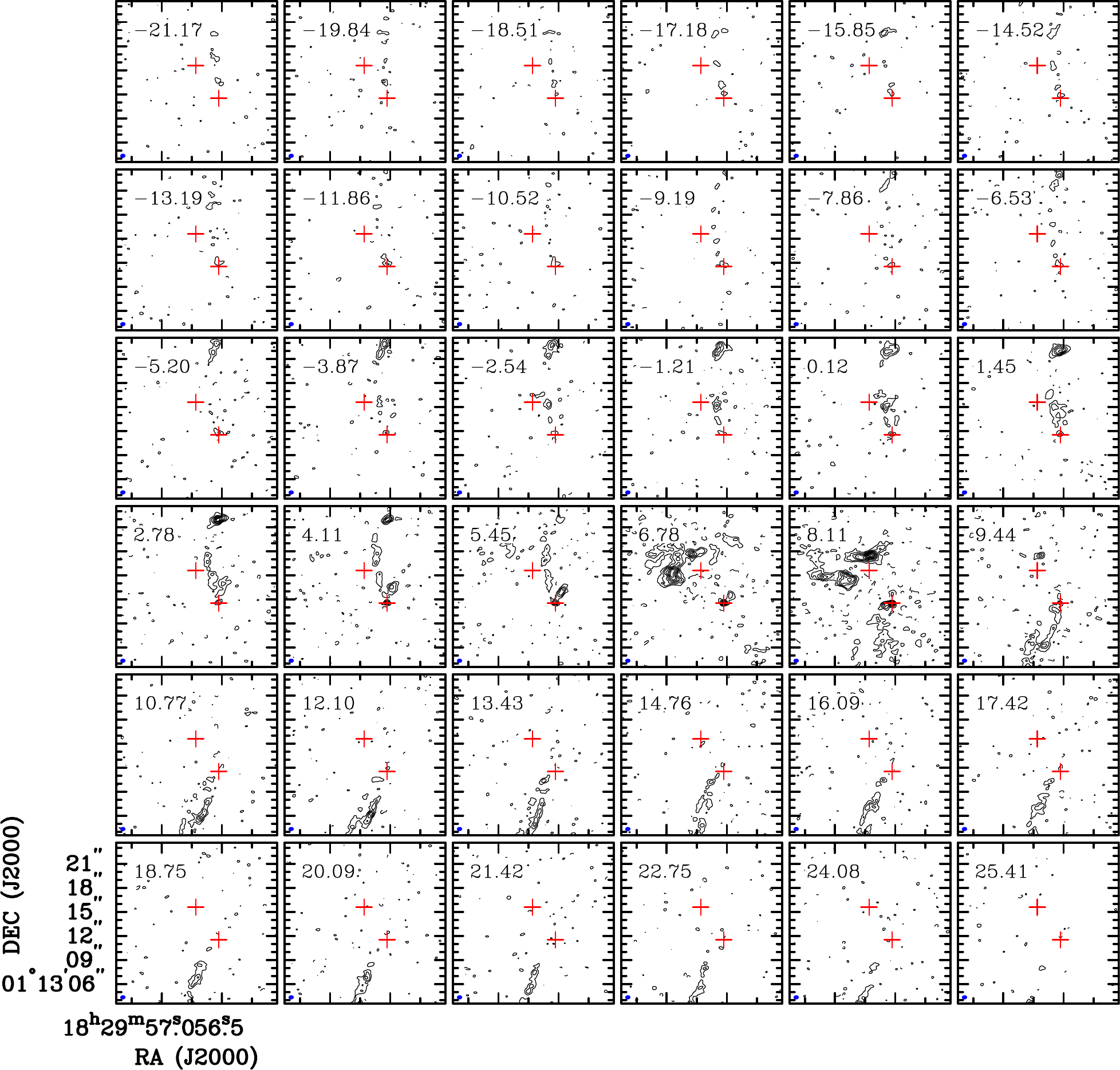}
\caption{Channel maps of the SO $J_N=6_5-5_4$ emission line. The velocity resolution is $1.33\ \kms$. Contour levels are from $3\sigma$ in steps of $3\sigma$, where $1\sigma$ corresponds to $4.0\ \mJB$. Two red plus signs and a blue ellipse in each panel are the continuum peak positions and the ALMA synthesized beam, $0\farcs 65\times 0\farcs 52,\ {\rm P.A.}=-85^{\circ}$, respectively.
\label{fig:soch}}
\end{figure}

\begin{figure}[ht!]
\epsscale{1.2}
\plotone{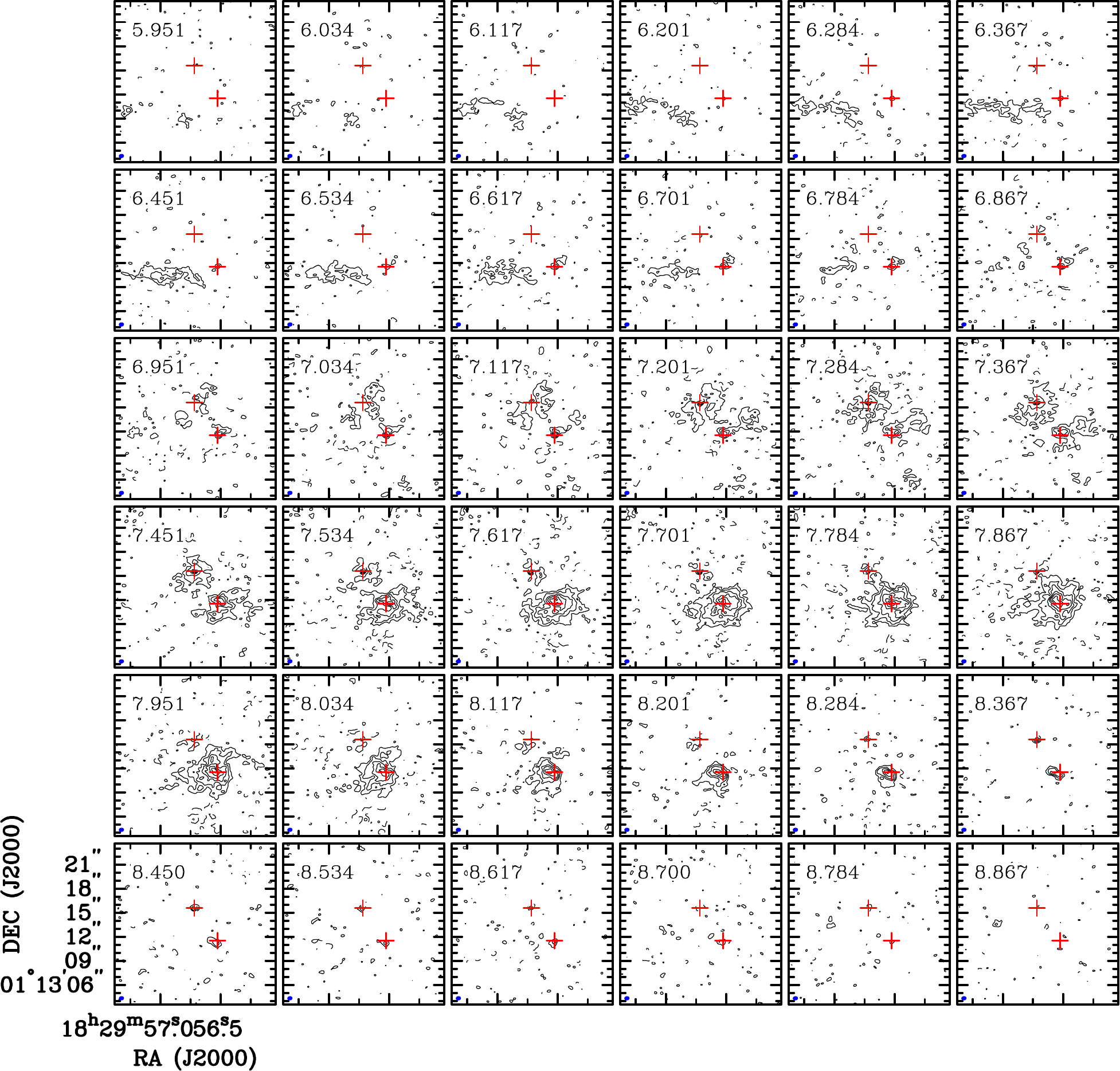}
\caption{Channel maps of the C$^{18}$O $J=2-1$ emission line. The velocity resolution is $0.083\ \kms$. Contour levels are from $3\sigma$ in steps of $3\sigma$, where $1\sigma$ corresponds to $12\ \mJB$. Two red plus signs and a blue ellipse in each panel are the continuum peak positions and the ALMA synthesized beam, $0\farcs 64\times 0\farcs 52,\ {\rm P.A.}=-83^{\circ}$, respectively.
\label{fig:c18och}}
\end{figure}

\bibliographystyle{aasjournal}
\bibliography{reference_aso}

\end{document}